\documentclass{emulateapj}

\pdfoutput=1

\usepackage{graphicx}

\usepackage[z,alpha]{tmsfnts}

\def\***#1{{\boldmath\textbf{\textsf{***#1***}}}}

\def\mum{\,\ensuremath{\mu\rm m}}
\def\vmax{\ensuremath{v_{\rm max}}}
\def\Vmin{\ensuremath{V_{\rm min}}}
\def\Mtot{\ensuremath{M_{\rm tot}}}
\def\Mvir{\ensuremath{M_{\rm vir}}}
\def\Msun{\ensuremath{M_{\odot}}}
\def\sun{\ensuremath{_{\odot}}}
\def\kms{km~s$^{-1}$}

\newcommand\f[1]{\ensuremath{S_{#1\mum}}}
\begin{document}

\title{Clustering of star-forming galaxies detected in mid-infrared with the \emph{Spitzer} wide-area survey}

\author{S.~Starikova\altaffilmark{1,2},
S.~Berta\altaffilmark{3},
A.~Franceschini\altaffilmark{2},
L.~Marchetti\altaffilmark{2},
G.~Rodighiero\altaffilmark{2},
M.~Vaccari\altaffilmark{2,4},
 and A.~Vikhlinin\altaffilmark{1,5}
 }
\altaffiltext{1}{Harvard-Smithsonian Center for Astrophysics, 
                  60 Garden Street, Cambridge, MA 02138, USA}
\altaffiltext{2}{Dipartimento di Astronomia, Universit\`a di
   Padova,Vicolo dell'Osservatorio 3, 35122 Padova, Italy}
\altaffiltext{3}{Max-Planck-Institut f\"ur Extraterrestrische Physik (MPE), 
                   Postfach 1312, 85741 Garching, Germany}
\altaffiltext{4}{Astrophysics Group, Physics Department, University of the Western Cape,
Private Bag X17, 7535, Bellville, Cape Town, South Africa}
\altaffiltext{5}{Space Research Institute (IKI), Profsoyuznaya 84/32,
                 Moscow, Russia}

\slugcomment{The Astrophysical Journal, in press}
\shortauthors{STARIKOVA ET AL.}
\shorttitle{CLUSTERING ANALYSIS OF 24\mum{} SOURCES}

\begin{abstract}
We discuss the clustering properties of galaxies with signs of ongoing star formation detected by the \emph{Spitzer Space Telescope} at $24\mum$ band in the SWIRE Lockman Hole field.
The sample of mid-IR-selected galaxies
includes $\sim20000$ objects detected above a flux threshold of
$\f{24}=310\,\mu$Jy. We adopt optical/near-IR color selection
criteria to split the sample into the lower-redshift and higher-redshift galaxy
populations. We measure the angular correlation function on scales of
$\theta=0.01-3.5$\,deg, from which, using the Limber inversion along with the redshift distribution established for similarly selected source populations in the GOODS fields \citep{Rodighiero2009}, we obtain
comoving correlation lengths of $r_0 = 4.98 \pm 0.28 \, h^{-1}\,$Mpc
and $r_0 =8.04 \pm 0.69 \, h^{-1}\,$Mpc for the low-$z$ ($z_{\rm
  mean}=0.7$) and high-$z$ ($z_{\rm mean}=1.7$)
subsamples, respectively. Comparing these measurements with the
correlation functions of dark matter halos identified in the
\textit{Bolshoi} cosmological simulation \citep{Klypin2010}, we find
that the high-redshift objects reside in progressively more massive
halos reaching $\Mtot\gtrsim3\times10^{12}\,h^{-1}\,\Msun$, compared to
$\Mtot\gtrsim7\times10^{11}\,h^{-1}\,\Msun$ for the low-redshift population.
Approximate estimates of the IR luminosities based on the
catalogs of $24\mum$ sources in the GOODS fields
 show that our high-$z$ subsample represents a
population of ``distant ULIRGs"
with $L_{\rm IR}>10^{12}L_{\odot}$, while the low-$z$ subsample mainly
consists of ``LIRGs'', $L_{\rm IR}\sim10^{11}L_{\odot}$.  The
comparison of number density of the $24\mum$ selected galaxies and of
dark matter halos with derived minimum mass $M_{\rm tot}$ shows that only 20\% of such halos may host
star-forming galaxies.
\end{abstract}

%-------------------------------------
\section{Introduction}
\label{sec:introduction}

The cosmic infrared background \citep[CIB;][]{1996A&A...308L...5P,
  1998ApJ...508...25H} accounts for approximately half of the total
extragalactic background energy integrated over cosmic time and
wavelengths \citep[e.g.,][]{2006A&A...451..417D,
  2001ARA&A..39..249H}. The CIB emission is mainly contributed by
star-forming galaxies where optical--UV light from young stellar
populations is absorbed by dust and re-emitted at longer wavelengths.
The IR-energy output per unit volume must strongly increase with
redshift to account for the total measured CIB
\citep{2001ARA&A..39..249H, 2005ARA&A..43..727L}. Indeed, observations
with the \emph{Infrared Space Observatory}
\citep[\emph{ISO};][]{2000ARA&A..38..761G} and the \textit{Spitzer Space
Telescope} \citep{2004ApJS..154....1W} revealed large number of distant
mid- and far-infrared sources \citep{2001ApJ...556..562C,
  2002A&A...384..848E, 2005ApJ...632..169L}. According to the current
consensus from both theoretical and observational studies, major
developments in the evolution of galaxies in the universe happened at
high redshifts, $z>1$ \citep[for references and details,
see][]{Franceschini2010}, with the peak of star formation and nuclear
activity occurring at $z\sim2$ \citep[e.g.,][]{1996MNRAS.283.1388M,
  2004ApJ...615..209H, 2005ApJ...624..630S,2011ApJ...737...90B}. A
large fraction of energy emitted during these active phases of galaxy
evolution is hidden by dust and can be detected only through mid- and
far-IR observations. Therefore, studying the distant universe in the
infrared provides valuable information on the history of assembly of
present-day massive galaxies \citep[e.g.,][]{2008ARA&A..46..201S, 2009ApJ...703..222L, 2009A&A...504..751S, Franceschini2010}.

In this work, we use observations of star-forming galaxies made by the
\textit{Spitzer Space Telescope} at $24\mum$. The \textit{Spitzer}
$24\mum$ surveys have revolutionized studies of ``distant
ULIRGs'' --- ultraluminous infrared galaxies. These
objects are dusty star-forming galaxies with infrared luminosity
$L_{\rm IR}>10^{12}L_{\odot}$\footnote{$L_{\rm
    IR}=\int_{8\mum}^{1000\mum}L_{\lambda}d\lambda$,
  \citep{Sanders1996}}
\citep[e.g.,][]{2004ApJS..154..160R,2005ApJ...628..604Y,2007ApJ...670..156D,2010A&A...524A..33F,2010ApJ...719..425F}. While
the average spectral energy distribution of high-$z$ sources is
consistent with that of present-day ULIRGs, the nature and the
cosmological environment hosting them must still be clarified
%\footnote{Most local ULIRGs have disturbed morphologies, consistent with being merging systems.} 
\citep[see][for details and references]{Huang2009}. Various photometric techniques are applied to identify high-redshift objects among the thousands detected by wide-field \textit{Spitzer} surveys, e.g.,~\cite{2005ApJ...628..604Y}, \cite{2007MNRAS.375.1121M}, \cite{2008ApJ...677..957F}, \cite{2009ApJ...692..422L}, \cite{2009A&A...508..117F}, \cite{Huang2009}, and \cite{2008ApJ...677..943D}. All these selected objects represent sub-populations of ULIRGs with observational characteristics partly overlapping those of star-forming galaxies detected in optical and submillimeter \citep[see recent papers by][]{Huang2009, 2009A&A...508..117F}. The nature  of these populations has been a subject of intensive work based on modeling of their physical properties such as spectral energy distribution (SED), star formation rate, stellar and halo masses, etc. \citep[e.g.,][]{2004ApJ...600..580G,2010MNRAS.404.1355D,2010MNRAS.407.1701N,2010MNRAS.405....2L}. A significant new observational input for such studies can be provided by measurements of the clustering amplitude, which is a unique tool for determination of the halo masses of high-redshift galaxies. The goal of this paper is to present clustering and halo occupation analysis of $24\mum$ detected galaxies from 
one of the largest \textit{Spitzer} extragalactic survey. 

First studies on clustering of $24\mum${} galaxies were made either in small fields with low statistics,
e.g., ~\cite{Gilli2007} and \cite{Magliocchetti2008}, or applying additional selection criteria as in \cite{Farrah2006} and \cite{Brodwin2008}. Here we improve on these first measurements by using a large sample of $\sim20,000$ galaxies detected in the Lockman Hole field, $\sim 8$~deg$^{2}$, and uniformly selected only by their $24\mum$ flux, $\f{24}>310\,\mu$Jy. 
Our data reduction procedures are presented in Section \ref{sec:data}. The clustering strength measurements of $24\mum$ selected galaxies and inferred properties of their dark matter (DM) halos are discussed in Sections \ref{sec:24mum_wtheta} and \ref{sec:24mum_halo}. Comparison with previously published results is presented in Section \ref{sec:24mum_comparison},
and our conclusions appear in Section 6.

Throughout the paper, all cosmology-dependent
quantities are computed assuming a spatially flat model with
parameters $\Omega_{\rm M}=0.268$ and $\Omega_{\rm \Lambda}=0.732$ \citep[best-fit
$\Lambda$CDM parameters obtained from the combination of CMB,
supernovae, baryon acoustic oscillations, and galaxy cluster data, see][]{2009ApJ...692.1060V}.
All distances are comoving and given with explicit \textit{h}-scaling, where the Hubble constant
is $H_0=100\,h^{-1}\,\rm{km\,s^{-1}\,Mpc^{-1}}$. The parameter uncertainties are quoted at a confidence level of 68\%.
IR luminosities were computed using $H_0=70\,\rm{km\,s^{-1}\,Mpc^{-1}}$ \citep[see][\,for details]{Rodighiero2009}.

%--------------------------------------------------------------------------------------------------------------------------------------------------------------------------------------
\section{The data sample}                     
\label{sec:data}
%--------------------------------------------------------------------------------------------------------------------------------------------------------------------------------------

For reliable clustering measurements one needs a statistically complete, large, and homogeneous sample of sources selected over a large area of the sky to probe the correlation signal on a wide range of scales. The Spitzer Wide-area InfraRed Extragalactic Survey \citep[SWIRE,][]{Lonsdale2003} is highly suitable for this purpose, as was demonstrated in several papers \citep{Waddington2007, Torre2007, Farrah2006}.
It is the largest survey carried out with the \emph{Spitzer Space Telescope}, covering $\sim 49\,\rm{deg}^2$ in six separate fields in the Northern and Southern sky. Each field was imaged in the seven near-to-far infrared bands:
InfraRed Array Camera (IRAC) 3.6, 4.5, 5.8, 8.0\,$\mum$ \citep{Fazio2004} and Multiband Imaging Photometer for Spitzer (MIPS) 24,
70, and 160\,$\mum$ bands \citep{Rieke2004}. In addition to the infrared observations, every SWIRE field has high-quality ancillary data. 

Following the goal of our work to estimate the correlation function of
star-forming galaxies detected in the MIPS $24\,\mum$ band, we first
selected a sample of bright sources, $\f{24}>400\,\mu$Jy, from the
SWIRE ELAIS-S1 catalog (M. Vaccari et al., in preparation). However, our
estimated angular correlation function, $w(\theta)$, showed an
unexpected lack of clustering signal at scales $\theta<36''$. There
were suggestions in the literature \citep[e.g.,][]{Gilli2007} that
because of the poor angular resolution of the MIPS instrument ($\sim
6''$ FWHM), there could be difficulties in determining $w(\theta)$ for
\emph{faint} sources due to blending. However, the deficit of close
pairs in the sample of \emph{bright} sources remained
unexplained. This problem has no bearing on our main results presented
below but obviously its origin needs to be understood. To this end, we
carried out a comparison of the angular correlation function of the
$24\,\mum$ sources selected from the four largest SWIRE fields
(Lockman Hole, ELAIS-N1, ELAIS-N2, and CDFS) using two releases of the
SWIRE team catalogs (versions 2005 and 2010), and an additional source
catalog based on the wavelet decomposition algorithm
(Section\,\ref{subsec:wavelet}). This comparison is reported in the
Appendix. Our clustering results for 24\,\mum{} sources presented
below are based on the best available catalog in the Lockman Hole
field.

\subsection{Wavelet-based Detection of 24\mum{} Sources}
\label{subsec:wavelet}
\def\wvdecomp{\textsc{wvdecomp}}

Due to the reasons outlined in the Appendix, we perform clustering
analysis of 24\mum{} sources extracted from the publicly available
MIPS images using the wavelet decomposition source detection algorithm
\citep[\wvdecomp, see][]{Vikhlinin1998}. This algorithm at
$\f{24}\gtrsim 300\,\mu$Jy performs nearly identically to the
detection method used in the Final SWIRE Data Release (J. A. Surace et al.,
in preparation) in terms of the $\log N - \log S$ distribution of
detected sources and their angular correlation function at large
scales. The only noticeable difference is in the treatment of very
crowded regions and zones in the immediate vicinity of the bright
sources (see the Appendix). These differences have no effect on our
clustering results presented in Section 3 and~4 below.

\wvdecomp{} was designed to efficiently detect both point-like and
slightly extended sources in the crowded fields. Originally, the
wavelet decomposition program was intended for Poisson-noise-limited
X-ray images, where it generally outperforms its rivals
\citep{2007A&A...473..857R}, but it was found that with a suitable
choice of parameters, it produces good results also for the 24\,\mum{}
MIPS images.

First, we re-bin the archival MIPS images to $2.4''$ pixels (by a
factor of two with respect to an original pixel size of $1.2''$) to
reduce the cross-correlation of noise in the adjacent pixels while
still maintaining the adequate sampling of the PSF. We then convolve
the image with the ${\rm scale}=2$ wavelet filter, corresponding to an
effective kernel width of $\approx 5''-6''$, matching the size of the
MIPS 24\,\mum{} point sources. The rms of variations in this convolved
image, excluding the regions around bright sources using
$\sigma$-clipping, is the approximation of effective noise at the
scale we are most interested in. This noise level is supplied to the
\wvdecomp\ program (its internal noise determination algorithm is best
suitable for the case Poisson statistics and thus not applicable for
MIPS images). \wvdecomp\ starts with the smallest scales and
iteratively detects and removes detected structures from the input
image, while adding them to the resulting ``clean'' image. When the
process is finished at the given scale, it proceeds to the next at
which the size of the wavelet kernel is increased by a factor of~two. In
our case, the detection algorithm works on the scales corresponding to
structure sizes (FWHM) of $\approx 2.4''$, $5''$, and $10''$,
bracketing the range of sizes for the MIPS point sources. Detection
threshold is set at $4.5\sigma$, at which we expect $\sim 100$ false
detections in the Lockman Hole area.\footnote{The calibration of the false
  detection rate was described in \citet{1995ApJ...451..553V}, and was done assuming uncorrelated Gaussian or Poisson
  noise in the image pixels. The noise properties in the SWIRE images
  are more complex but the above value is still a good
  order-of-magnitude estimate of the false-positive rate in our
  24\,\mum{} sample.}

The main output of the wavelet decomposition algorithm is a list of
source locations detected above a predefined SNR threshold, and a map
which allows one to split the original image into ``empty'' regions
and those with significant emission ``belonging'' to a particular
source. The source fluxes were then measured using aperture
photometry. In choosing the aperture size, the tradeoff is between our
desire to include as much of the source flux as possible into the
aperture size, and the fact that for wide apertures, the flux
measurements are increasingly affected by the larger-scale background
fluctuations and by source confusion. Several tests have shown that
the best results are achieved for an aperture size of $4''$,
encompassing approximately 50\% of the PSF power, and corresponding to
the bright core of the MIPS PSF. These aperture fluxes were then
converted into total flux using the PSF model calibrated with images
of the bright stars in the same field. Using this method, the
24\,\mum{} sources were extracted from the MIPS map of the Lockman
Hole field.

\subsection{The Lockman Hole Source Sample}
\label{subsec:sample}

The Lockman Hole is the largest of the SWIRE fields. In addition to
deeper MIPS observations (the limiting flux is $\f{24}=310 \,\mu$Jy,
compared, e.g., to $\f{24}=400 \,\mu$Jy in the ELAIS-S1 field, see
Appendix~\ref{sec:flux_limits} for details), it has deep and uniform
data in many other bands. In particular, we used the data from the
Two Micron All Sky Survey (2MASS) survey for the star-mask construction (see
Section~\ref{subsubsec:mask_construction}), and the optical observations
carried out with INT-WFC and KPNO MOSAIC1 \citep{2011MNRAS.416..927G}
to photometrically separate the 24\,\mum-selected objects into the
low- and high-redshift subsamples
(Section~\ref{subsubsec:24mum_color_mag}).

We cross-correlated our sample of 24\,\mum{} sources with the
multi-band IRAC-based catalog (limiting fluxes of $\f{3.6}\simeq7
\,\mu$Jy and $\f{4.5}\simeq11 \,\mu$Jy, M. Vaccari et al., in
preparation) using a matching radius of $3.2''$. We then applied the
following flux cuts: $310<\f{24}<2500 \,\mu$Jy and $\f{3.6}<1000
\,\mu$Jy, and $\f{4.5}<1000 \,\mu$Jy. $\f{24}=310 \,\mu$Jy is the flux
at which the catalog is complete and the fluxes are measured reliably
and accurately. The bright flux cuts are applied in order to
conservatively discard obviously extended and/or saturated sources
whose astrometry may be poor and whose flux estimates may be affected
by saturation. Only $1.7\%$ of sources with $\f{24}>310\,\mu$Jy had no
IRAC-couterparts.  A small fraction of them are Galactic stars, $\sim
0.3\%$ are expected due to false detections for our choice of
\wvdecomp{} detection thresholds, the nature of the rest is unclear.  In any
case, their number is too small to affect our clustering measurements.

\subsubsection{Elimination of Stars and the Region Mask}
%\subsubsection{Removal of stellar contamination  \***{Elimination of stars, Star-galaxy separation and stellar mask}}     
\label{subsubsec:star_gal_sep}
\label{subsubsec:mask_construction}

Galactic stars contaminate our clustering 
analysis of extragalactic sources and should be removed.\footnote{We note, however, that the star removal
is not a crucial component of our analysis since the contamination
of near- to mid-IR galaxy samples by foreground stars is a severe
problem only at fluxes of brighter than several mJy.}
To this end, we followed the procedures of ~\cite{Shupe2008} and ~\cite{Waddington2007} in which the foreground stars were
identified using the 2MASS Point Source Catalog \citep{Skrutskie2006}. 
The derived $24 \mum$-IRAC catalog was cross correlated with the 2MASS
survey using a matching radius of $2.5''$.
 \cite{Shupe2008} proposed that 
nearly all of the $24\,\mum$-emitting sources with color $K_{s}-[24]<2.0$ (Vega, mag)
are Galactic stars (see their Figure \,2). We applied this criterion to our
catalog and eliminated such sources.

In addition to directly polluting the extragalactic sample, bright
Galactic stars may affect our clustering measurements indirectly, by
obscuring the background galaxies or affecting the fluxes of the
fainter galaxies near the same line of sight. Therefore, we need to
completely exclude from the analysis the sky regions affected by the presence of 
bright foreground stars. Following \cite{Waddington2007} this was achieved by masking out the circular
regions around sources with $K_s<12$~(Vega, mag) from the cross-correlated
$24\mum$-IRAC-2MASS catalog; the exclusion radius was determined
as $\log(R'')=3.1-0.16\,K_s$, which is the distance at which the
stellar PSF merges into the background \citep{Waddington2007}. 

A close examination of the $24\mum$ source catalog shows that
there are spurious detections around very bright $24\mum$
sources (most of which correspond to Galactic stars or low-$z$
galaxies). Therefore, we decided to mask out those regions as well.
The exclusion radius was set to be $20''-80''$, depending on the source flux.
%In the region mask (%Figure~\ref{fig:24mum_mask_lh}) the circles mark 
%locations of bright stars and extremely bright 24\,\mum{} sources.

\begin{figure}
{\centering
\includegraphics[width=0.85\linewidth]{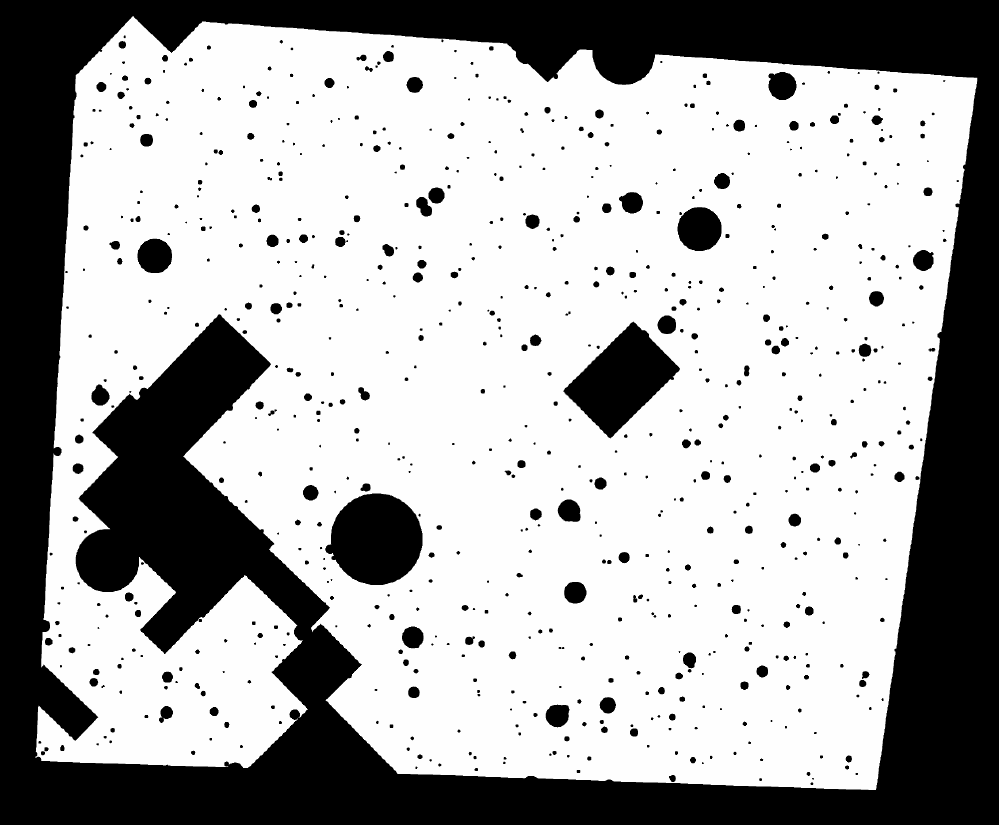}
\par}
\caption{Final region mask for the clustering analysis in the Lockman
Hole field. The circles mark the locations of stars and bright
  objects. The rectangles mask those regions where the completeness of
  INT/WFC images is not achieved for $i=22.8$ (AB mag). All black
  patches were excluded from the subsequent analysis (see details in Section~\ref{subsubsec:mask_construction}).}
\label{fig:24mum_mask_lh}
\end{figure}

As we will discuss in the next section,
Section~\ref{subsubsec:24mum_color_mag}, we use the INT/WFC optical data to
divide our sample photometrically into the low- and high-redshift
subsamples. Unfortunately, the INT/WFC observations are insufficiently
deep in some subsections of the MIPS Lockman Hole image, and we had to
mask out those regions also.  To identify the regions of insufficient
INT/WFC depth, we examined the distribution of optical counterparts
for $3.6\,\mum$ IRAC sources at various $i$-band magnitude cuts. We
found that the depth is at least $i=22.8$ throughout the field, except
for the regions masked out as rectangles in
Figure~\ref{fig:24mum_mask_lh}. At fainter magnitudes, the WFC coverage
becomes highly nonuniform. %Therefore, we use a limit of
%$i=22.8$ for the redshift separation
%(\S~\ref{subsubsec:24mum_color_mag}), and, the final region mask shown in
%%Figure~\ref{fig:24mum_mask_lh} for the estimation of the angular
%correlation function.  

The resulting mask excluding the regions around bright stars,
extremely bright 24\,\mum{} sources and the regions of nonuniform
optical coverage is shown in Figure~\ref{fig:24mum_mask_lh}, and was
used in the estimation of the angular correlation function
(Section~\ref{sec:24mum_wtheta}). The total ``good'' survey area is
7.9~deg$^{2}$.

\subsubsection{Identifying Low- and High-redshift Galaxy Populations}
\label{subsubsec:24mum_color_mag}

To derive the spatial correlation length and investigate the dependence of
clustering on redshift, we need to know the redshift distribution of
the sources. Unfortunately, the vast majority of the 24\mum{}
sources selected in the Lockman Hole field have neither spectroscopic
nor photometric redshifts.  The SWIRE photometric redshift catalog
\citep{photz2008}, available in this field, has a limited and heavily
inhomogeneous coverage for our sample. The approach we are taking
instead is to use simple photometric criteria to divide the
catalog into the low- and high-redshift subsamples, and then use
a similarly selected sample of $24\mum$ sources from the GOODS survey
to derive the redshift distribution within each subsample.

To separate the sample into low- and high-redshift sources, we defined
the optical-to-NIR color selection criterion based on the optical $I$-band data \citep[from
ESIS-VIMOS survey;][]{Berta2008} and SWIRE IRAC $4.5\,\mum$ observations in the 
ELAIS-S1 SWIRE field. Particularly, we examined the dependence of the $(I-[4.5])_{\rm AB}$ 
color on redshift for various galaxy spectral templates such as Mrk~231 (Sy-1), IRAS~19254 (Sy-2),
M~82 (starburst), M~51 (spiral), and NGC~4490 (blue spiral) \citep[see examples of a similar analysis
in][]{Berta2007, Berta2008}. It appears that for starburst galaxies, the color cut $(I-[4.5])_{\rm AB}\sim 3$ separates well low
($z\lesssim1$) and high ($z\gtrsim1$) redshift galaxy populations, with only a small contamination in both groups.
Such a rapid color transition around $z\sim1$ can be explained by the passage of the Balmer break
in the galaxy spectra through or redward the $I$ band.
%\***{The color $(I_{\rm VIMOS}-[4.5])_{\rm AB}$ versus IRAC $4.5\,\mum$-band magnitude diagram plotted for the $24\,\mum$ sources with
%$\f{24}>400\,\mu$Jy detected in the SWIRE + ESIS-VIMOS ELAIS-S1 field along with the redshift tracks for the spectral templates of galaxies (mentioned above) additionally confirmed that a
%color cut at $(I_{\rm VIMOS}-[4.5])_{\rm AB}\sim 3$ can be used to
%divide the observed sources into low- and high-$z$ galaxy populations.}

To further refine this color selection criterion,
we applied it to the deep \emph{Spitzer} observations of GOODS fields
\citep{Rodighiero2009}. The GOODS-N and GOODS-S $24\,\mum$ catalogs
include 889 and 614 sources, respectively, detected in a total area of
$\sim 350\,\text{arcmin$^2$}$. The catalogs are complete down to
$\f{24}=80\,\mu$Jy. Observations in the \textit{i}~band were made by
the Advanced Camera for Surveys in both fields down to a magnitude limit \textit{i}=26.5
\citep{Grazian2006}. Redshift estimates are available for
all these sources, 46\% are spectroscopic and 54\% photometric redshifts.
The latter are estimated with an rms scatter in
$z_{\text{phot}}-z_{\text{spec}}$ of 0.09 and 0.06 for the GOODS-N and
GOODS-S samples, respectively \citep[for details see][]{Rodighiero2009}.

\begin{figure*}
{\centering
\includegraphics[width=0.5\linewidth]{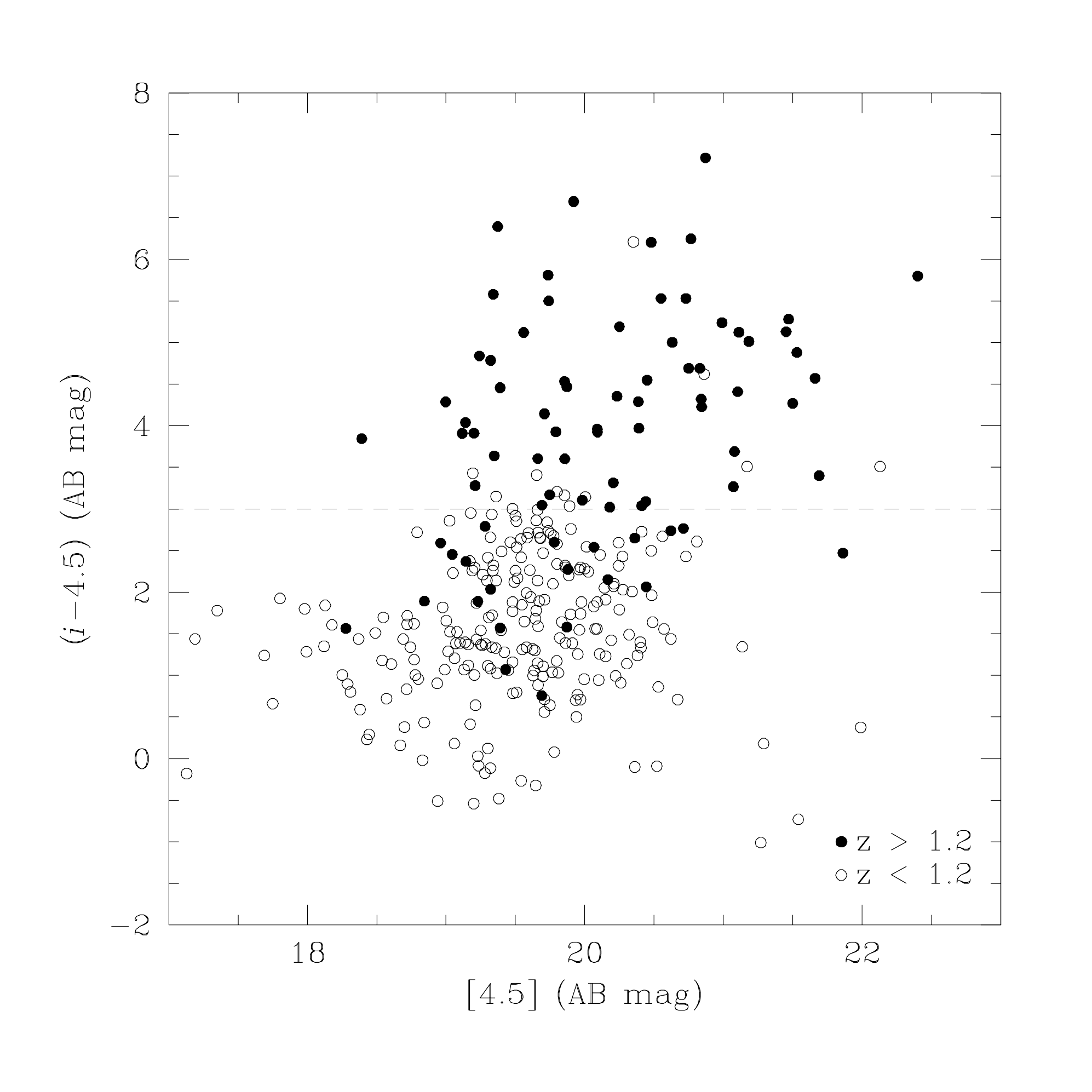}~
\includegraphics[width=0.5\linewidth]{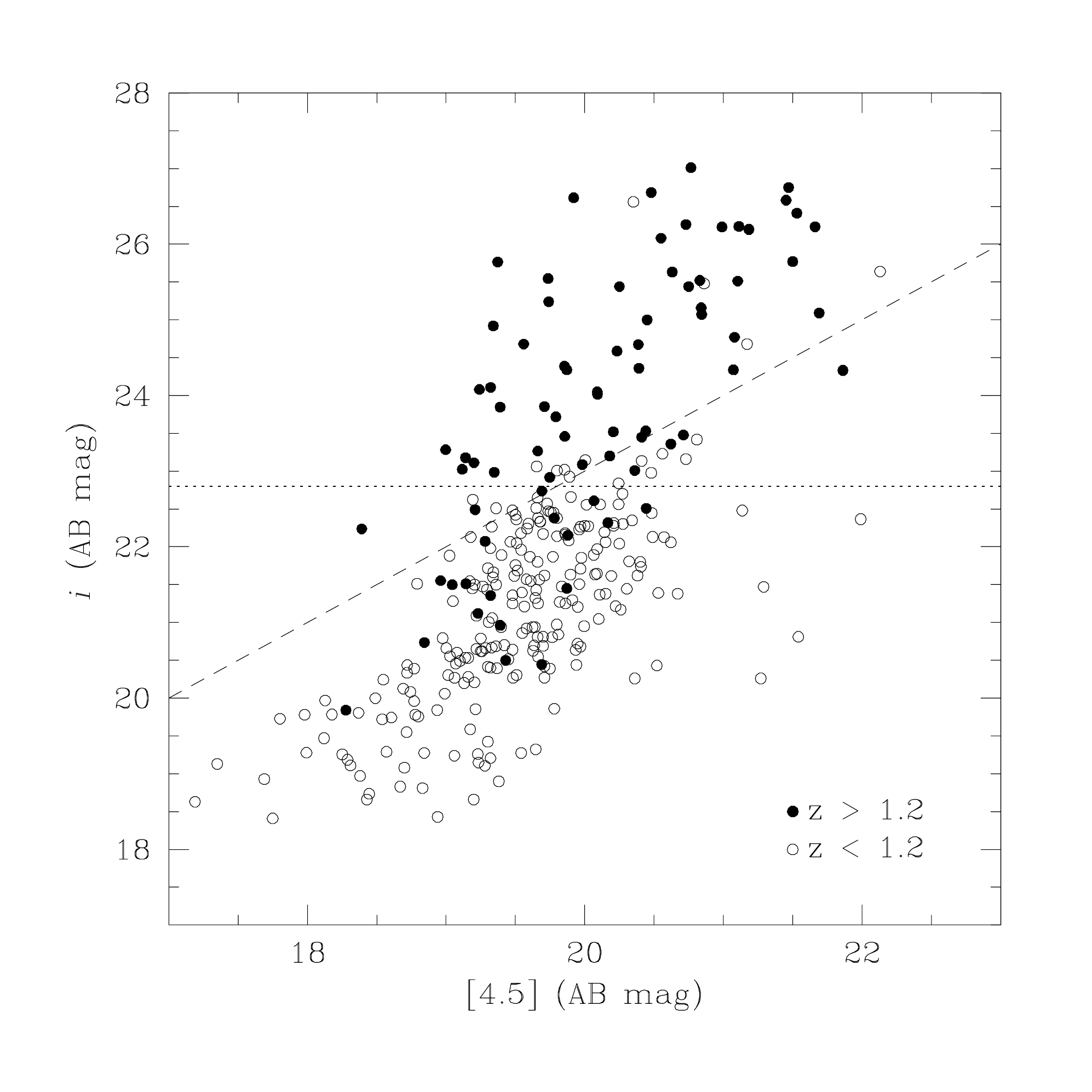}%
\par}
\caption{Color--magnitude (a) and magnitude $i$ vs.~magnitude [4.5] (b) diagrams for the GOODS-N and GOODS-S sources with $\f{24}>310 \,\mu$Jy. Open and filled circles are galaxies at redshifts lower and higher than $1.2$. On both figures, a dashed line represents $(i-4.5)=3$ (AB mag). A dotted line on the figure (b) corresponds to a magnitude $i=22.8$ at which the INT/WFC coverage is uniform in the Lockman Hole field.}
\label{fig:24mum_color}
\end{figure*}

From the GOODS catalogs, we selected the sources with
$\f{24}>310 \,\mu$Jy and separated them into two redshift bins
$z>1.2$ and $z<1.2$.\footnote{The boundary was chosen near the minimum
  of the bimodal redshift distribution predicted by the Franceschini
  et~al.~(\citeyear{Franceschini2010}) model.}  The color--magnitude diagram
for these sources shows that the low-
and high-$z$ galaxies indeed can be separated by a boundary value of
$(i-4.5)=3$ (AB mag) (dashed line in Figure~\ref{fig:24mum_color}(a)). The deepest optical data available
in the Lockman Hole field are those from the INT/WFC which provides
sufficiently uniform coverage to $i=22.8$ (with the $5\sigma$
magnitude limit reaching $i=23.3$ (AB) in the deepest sections of the
survey). Therefore, a magnitude cut of $i=22.8$ had to be
incorporated in our selection. Figure~\ref{fig:24mum_color}(b) shows
that the low-$z$ sources fainter than $i=22.8$ (above dotted line) and with the color $(i-4.5)<3$ (AB mag) (below dashed line)
in practice are very few and they only minimally
contaminate ($\sim10\%$) the high-$z$ sample. Based on these considerations, we
implemented the redshift separation as a combined color and magnitude
criterion: the source is considered to belong to a high-redshift
sample, if it is undetectable in the INT/WFC $i$ band, or its measured
$i$ magnitude is $>22.8$, or the $(i-4.5)$ (AB mag) color is $>3$.

One of the main sources of concern for the color--magnitude based
separation of 24\mum{} objects into low- and high-redshift subsamples
is the presence of active galactic nuclei (AGNs) in the sample.
%\***{The AGNs of both types do not follow the applied color selection 
%due to the shape of their SED, and thus can contaminate the high-redshift
%locus of color magnitude diagram.}
Therefore, we checked the
AGN contents in the GOODS sample of the 24\,\mum selected
sources. According to \cite{Rodighiero2009}, less than $10\%$ of these sources are type-1 AGNs. The authors
classified the observed SEDs using \cite{Polletta2007} templates. This AGN fraction is consistent
with that reported by \cite{Gilli2007} and \cite{Treister2006}, who used very deep \textit{Chandra} X-ray
observations in the GOODS fields. Concerning the highly obscured
(type-2) AGNs and the sources of composite spectral type
(starburst+ANG), their contribution to the $24\,\mum$ emitting sources
is hard to estimate. One of the reasons is that the AGN and
star formation activity often occur simultaneously, and both are
revealed in the form of the $24\,\mum$ emission \citep[see, e.g.,][and references therein]{Brand2009, Rodighiero2009, Franceschini2005}. Some studies suggest, on the basis of estimates
by different methods, that the $24\,\mum$ selected samples may contain
$\sim$20\%--30\% of AGNs of both types \citep{Sacchi2009,
  Franceschini2005}. However, we note that to estimate the redshift
distribution within our color and $i$-magnitude-selected subsamples,
we used an empirical redshift distribution of identically selected
GOODS sources (see below). As long as the GOODS redshifts are valid
and the GOODS sample is a fair representation of our main Lockman Hole
sample, the derived $dN/dz$ models for the low- and high-redshift
subsamples are correct, even though the high-$z$ subsample may be
slightly contaminated by AGNs.

\subsection{Empirical Redshift Distributions}
\label{subsec:24mum_redshift_distribution}

We need a model for the redshift distribution of the sources in order
to use the Limber equation (Equations~(\ref{frm:LE_power}) and (\ref{frm:LE_full}) below) to relate the angular and spatial
correlation functions. We determined these redshift distributions
empirically, using the GOODS sources selected identically to our main
sample in the Lockman Hole field. All sources with $\f{24}>310
\,\mu$Jy in GOODS-N and GOODS-S fields were divided into low- and
high-redshift subsamples by applying the color-magnitude selection
criteria (Section~\ref{subsubsec:24mum_color_mag} and Figure~\ref{fig:24mum_color}(b)). The obtained redshift distributions within
these photometrically-selected samples are shown in
Figure~\ref{fig:24mum_nz}(a) and (b). These empirical distributions can be
well approximated by a Gaussian model:

\begin{figure*}
{\centering
\includegraphics[width=0.5\linewidth]{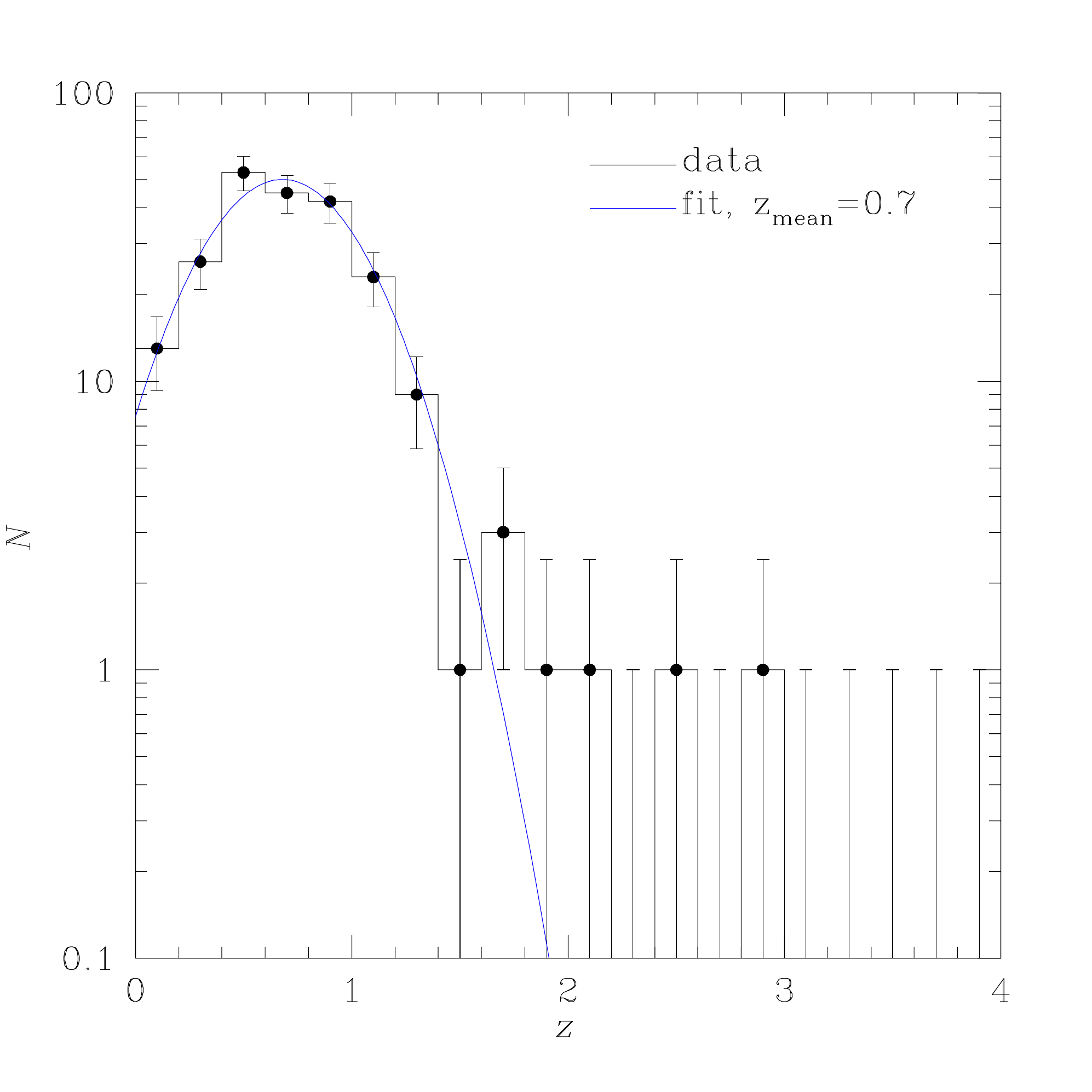}~
\includegraphics[width=0.5\linewidth]{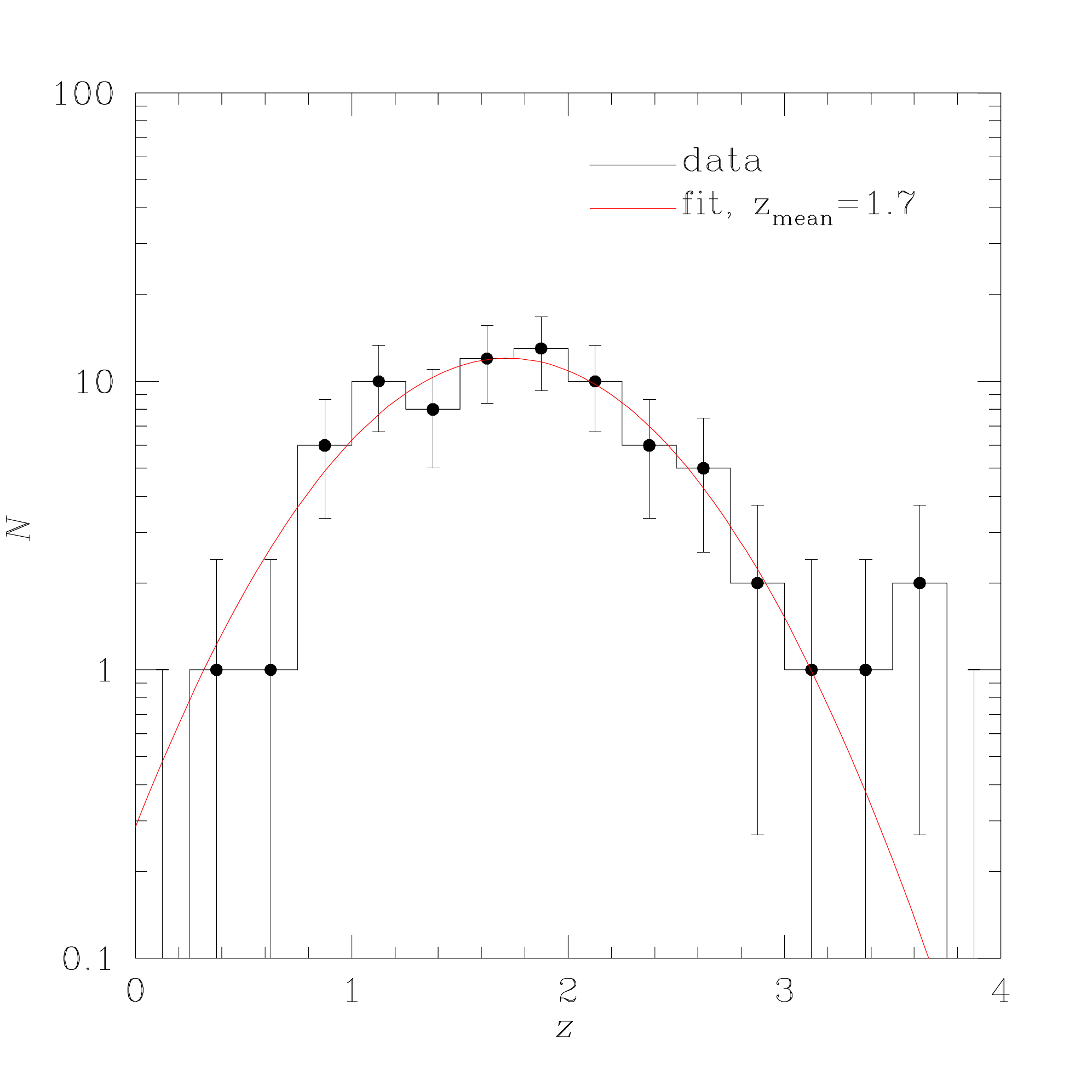}%
\par}
\caption{Redshift distribution of GOODS sources ($\f{24}>310 \,\mu$Jy) incorporated into low-$z$ (a) and high-$z$ (b) subsamples based on their color $(i-[4.5])$ and $i$-band magnitude. Blue and red lines are Gaussian fits with $z_{\rm mean}=0.7$ and  $z_{\rm mean}=1.7$, respectively.}
\label{fig:24mum_nz}
\end{figure*}

\begin{figure}
{\centering
\includegraphics[width=1.0\linewidth]{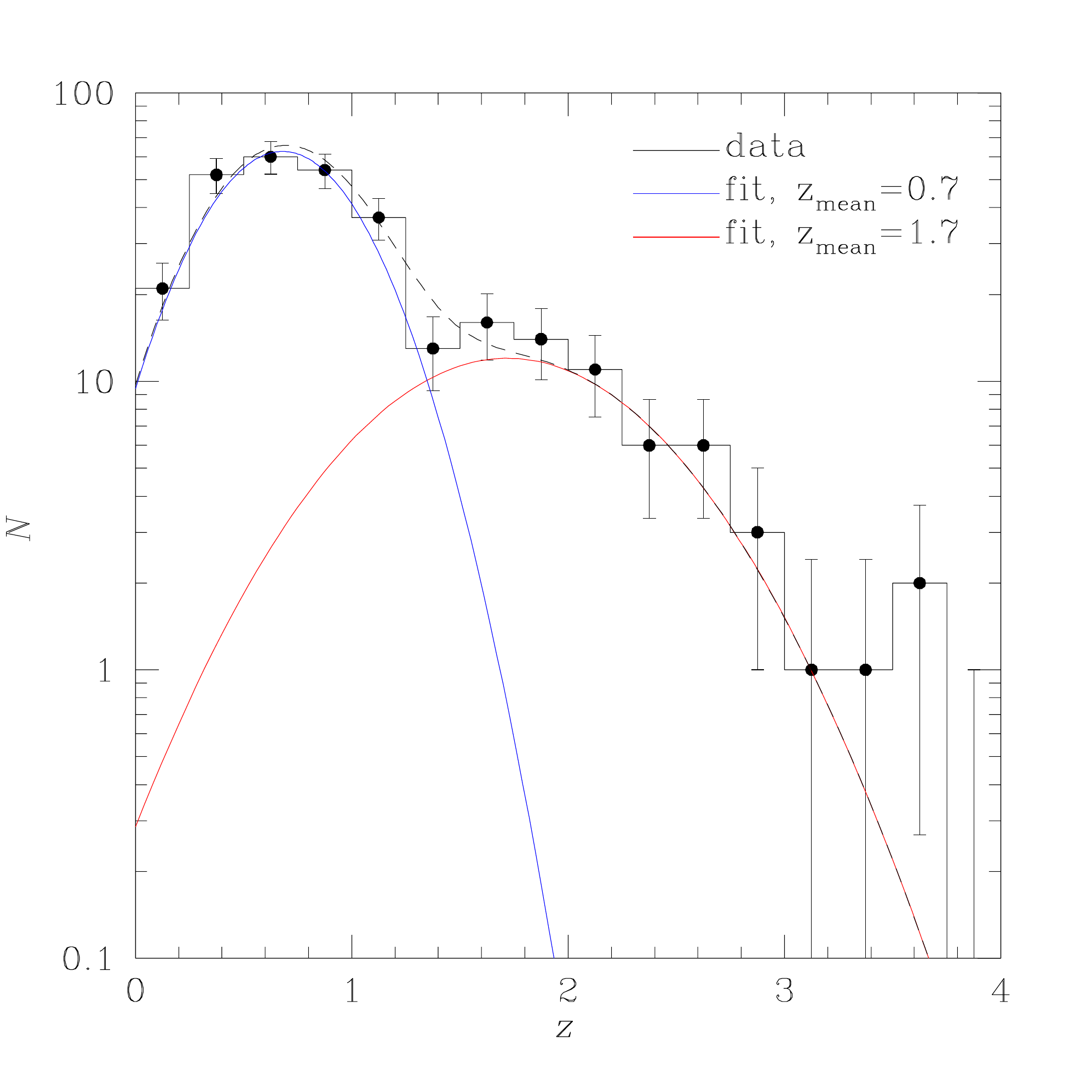}
\par}
\caption{Redshift distribution of sources brighter than $\f{24}=310 \,\mu$Jy from GOODS surveys.  Blue and red lines are Gaussian fits to redshift distributions of sources undergone color--magnitude selection. The dashed line is a combined fit of two selected samples.}
\label{fig:24mum_nz_all}
\end{figure}

\begin{equation}
  dN/dz = C\times\exp(-(z-z_{\text{mean}})^{2}/2\sigma^{2})
\end{equation}
(blue and red lines in Figure~\ref{fig:24mum_nz}). The best-fit
parameters for the low-$z$ subsample in the redshift range $0<z<2$ are
$C=50, \sigma=0.349, z_{\rm mean}=0.7$. For the high-$z$ subsample in
the redshift range $0.5<z<3.5$, we find $C=12, \sigma=0.629, z_{\rm
  mean}=1.7$. The derived widths are significantly larger than the
estimates uncertainties in the GOODS photometric redshifts ($\pm$0.06--0.09), and therefore accurately approximate the \emph{intrinsic}
widths of the redshift distributions for our two subsamples.

This two-Gaussian model provides a good fit also to the redshift
distribution of all GOODS sources with $\f{24}>310 \,\mu$Jy (i.e.,
without the photometric separation into low and high-$z$
subsamples). The combined redshift distribution is shown in
Figure~\ref{fig:24mum_nz_all}, and the dashed line is the sum of two
Gaussian models for the low and high-$z$ subsamples.

We also can use these subsamples of GOODS galaxies to estimate the
typical infrared luminosities (8\mum--1000\mum) for our Lockman
Hole sample.  In the GOODS low-redshift subsample, $z_{\rm mean}=0.7$,
the mean luminosity is $L_{\rm IR}\sim 3\times 10^{11}L_{\sun}$
indicating that the selected objects belong to the class of luminous
infrared galaxies \citep[``LIRGs'', $10^{11}L_{\sun}<L_{\rm
  IR}<10^{12}L_{\sun}$,][]{Sanders1996}.  The high redshift galaxies,
$z_{\rm mean}=1.7$, have an order of magnitude higher mean luminosity,
$L_{\rm IR}\sim 3\times 10^{12} L_{\sun}$ which places them into the
category of ultra-luminous infrared galaxies \citep[``distant
ULIRGs''; $L_{\rm IR}>10^{12}L_{\sun}$;][]{Sanders1996}.  Barring an
unexpectedly high level of cosmic variance, our $24 \mum$ sources
selected in the Lockman Hole field should have the same mean
luminosities.

%-----------------------------------------------------------------------------------------
\section{Clustering properties of $24\,\mum$ selected galaxies}
\label{sec:24mum_wtheta}
%------------------------------------------------------------------------------------------

The total area of the Lockman Hole field used in the clustering analysis (white regions in Figure~\ref{fig:24mum_mask_lh}) is $\simeq 7.9$~deg$^{2}$. There are
21844 \, $24\mum$ emitting objects with fluxes greater than
$310 \,\mu$Jy within this area. Applying the color--magnitude
selection criteria (Section~\ref{subsubsec:24mum_color_mag}), we obtained two subsamples
of 14822 and 7022 sources with $z_{\rm mean}=0.7$ and $z_{\rm
  mean}=1.7$, respectively.

\begin{figure}
{\centering
\includegraphics[width=1.0\linewidth]{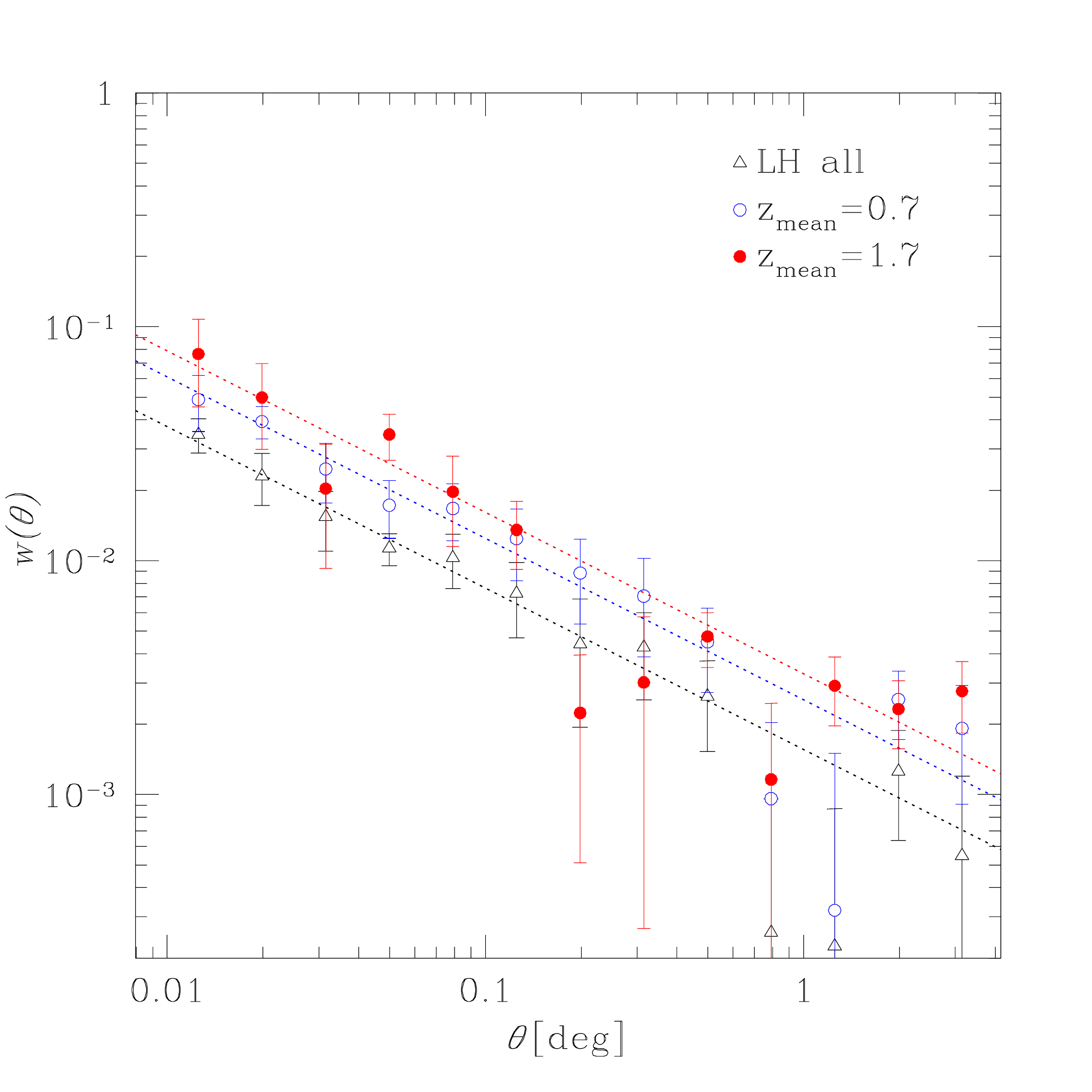}
\par}
\caption{Two-point angular correlation function of SWIRE Lockman Hole sources brighter than $\f{24}=310 \,\mu$Jy. The dotted lines are power-law fits. Triangles represent  clustering of the whole sample, open and filled circles are for the low-$z$ and high-$z$ galaxies, respectively.}
\label{fig:24mum_wtheta}
\end{figure}

The angular correlation functions were estimated by the Landy \&
Szalay method (\citeyear{1993ApJ...412...64L}) at angular scales
$0.01<\theta<3.5$ deg.\footnote{These angular sizes correspond to
  the comoving separations 0.12--43, 0.31--109, 0.50--174, and
  0.78-272 $h^{-1}\,$Mpc at $z=0.25$, $0.7$, 1.3, and 2.8,
  respectively (cf.\ Figure~\ref{fig:24mum_nz_all}).} The random
points used in this estimator were homogeneously distributed in the
field but avoiding the excluded regions of the mask shown in Figure~\ref{fig:24mum_mask_lh}. In order to suppress the uncertainties
related to a complex geometry of the field and to decrease the
statistical errors, the number of simulated random points was 100
times greater than the number of data points in each sample. The
correlation function was computed in angular bins
$\Delta\log\theta=0.2$. In Figure~\ref{fig:24mum_wtheta}, we show the
derived angular correlation functions for the whole sample (open black
triangles), for the low-$z$ subsample with $z_{\rm mean}=0.7$ (open
blue circles), and for high-$z$ subsample with $z_{\rm mean}=1.7$
(filled red circles).
  
%\begin{figure}
%{\centering
%\includegraphics[width=1.0\linewidth]{plots/w_all_jack_dotted.pdf}
%\par}
%\caption{Two-point angular correlation function of SWIRE Lockman Hole sources brighter than $\f{24}=310 \,\mu$Jy. The dotted lines are power-law fits. Triangles represent  clustering of the whole sample, open and filled circles are for the low-$z$ and high-$z$ galaxies, respectively.}
%\label{fig:24mum_wtheta}
%\end{figure}
    
Statistical uncertainties which can be assigned to angular correlation function $w(\theta)$ measured using
the Landy \& Szalay estimator are $\delta w(\theta)=1+w(\theta)/\!\!\sqrt{{\rm DD}(\theta)}$ 
\citep{1993ApJ...412...64L}, where DD is the number of data pairs.
However, it is considered that these uncertainties 
do not account for cosmic variance and covariance of the correlation
function at different separations, and therefore, underestimate real errors. 
These difficulties might be overcome by applying,
for instance, the jackknife subsampling of data \citep[e.g.,][]{2002ApJ...579...48S,
2002ApJ...571..172Z, Waddington2007, 2007MNRAS.381..573R}.
To calculate jackknife errors we divided the observed field into 25
approximately equal-sized patches and computed the correlation function 
excluding one part of our sample at one time. The ensemble errors are then
estimated from the scatter between perturbed and full sample realizations:
\begin{equation}
\sigma^2(\theta) = \sum_{i=1}^N\frac{{\rm DR}_i(\theta)}{{\rm DR}(\theta)}[w_i(\theta)-w(\theta)]^2,
\label{frm:jack_error}
\end{equation}
where DR is the number of pairs between cross-correlated data and random catalogs,
$i$ refers to a given sample realization, and ${\rm DR}_i/{\rm DR}$ accounts for a complex field
geometry \citep{2005MNRAS.359..741M, 2007MNRAS.381..573R}. 
All quoted uncertainties are obtained by applying the jackknife subsampling technique to the data, 
except in Appendix~\ref{sec:ang_fun_comparison}, where we compare the correlation functions 
from different catalogs and calculate errors $\delta w(\theta)$ (see above).   

Because of the good statistics of the SWIRE sample and the large size of
the Lockman Hole field, we are able to measure the clustering signal
at angular scales which correspond to fairly large spatial
scales. Indeed, comoving sizes of 1--8\,$h^{-1}\,$Mpc at $z=1.7$
correspond to an angular range of $0.017^{\circ}-0.13^{\circ}$. A
great advantage of the measurements done at such large scales is that
we directly probe the clustering signal at angular separations which
correspond to the expected range of three-dimensional correlation lengths,
$r_{0}$. This makes it possible to obtain robust
estimates of $r_{0}$ from a standard power-law fit to the angular
correlation function, $w(\theta)=(\theta/\theta_0)^{1-\gamma}$, and application of the simplified Limber equation
(full version is given by Equation~(\ref{frm:LE_full}))
which gives a direct link between the angular and spatial correlation lengths:
\begin{equation}
\theta_0^{\gamma-1} = r_0^{\gamma}\;A(\gamma){\displaystyle{\frac{2}{c}}}\;\frac{\int\limits_{0}^{\infty}dz\,N(z)^2\,H(z)D_{\rm M}(z)^{1-\gamma}}{\big[\int\limits_{0}^{\infty}dz\,N(z)\,\big]^2},
\label{frm:LE_power}
\end{equation}
where $D_{\rm M}(z)$ is the transverse comoving distance to redshift $z$ and $N(z)$ is the redshift distribution of sample galaxies.  $H(z)=H_0\sqrt{\Omega_{\rm M}(1+z)^3+\Omega_{k}(1+z)^2+\Omega_{\Lambda}}$ is the Hubble parameter at redshift $z$ and 
$A(\gamma)=\Gamma(1/2)\,\Gamma([\gamma-1]/2)/\Gamma(\gamma/2)$. 
If the angular
correlation function measurements at large scales are unavailable, a
power-law fit to the data at small angular/spatial scales may lead to
incorrect estimates of the correlation lengths and incorrect
conclusions about clustering properties of given galaxy populations
\citep[e.g.,][and references therein]{2004ApJ...609...35K, Quadri2007, Quadri2008}. 

The angular correlation functions shown in Figure~\ref{fig:24mum_wtheta} were iteratively fitted over the angular range
$0.01^{\circ}<\theta<3.5^{\circ}$ with a power-law model,
$w(\theta)=(\theta/\theta_0)^{1-\gamma} - {\rm IC}$, where the term IC refers to the Integral Constraint.
The IC correction accounts for a systematic offset in estimated correlation function due to the finite size
of any survey \citep{1974A&A....32..197P, peebles80} and it is usually calculated using a
method proposed by \cite{1993MNRAS.263..360R}:
\begin{equation}
{\rm IC}=\theta_0^{\gamma}\,\frac{\sum_j{\rm RR}(\theta_j)\,\theta_j^{1-\gamma}}{\sum_j{\rm RR}(\theta_j)},
\label{frm:IC}
\end{equation}
where ${\rm RR}(\theta_j)$ is the number of random pairs in an angular bin~$j$.

The best-fit parameters for the entire sample
are $\theta_0=0.31'' \pm 0.04''$, and $\gamma=1.69 \pm 0.11.$\footnote{
The uncertainties include the covariance of the parameters.}
Splitting
the whole sample into smaller subsamples obviously increases the
statistical uncertainties. Therefore, we decided to fix the power-law
slope in the subsequent analysis at $\gamma=1.69$. The best-fit
amplitudes for the low-$z$ and high-$z$ data are then $\theta_0=0.63''
\pm 0.09''$ and $\theta_0=0.91'' \pm 0.21''$,
respectively. %\footnote{For the high-$z$ sample, the fit was over the
%  angular range $0.01^{\circ}<\theta<0.13^{\circ}$ because at larger
 % angles, we expect that the correlation function can start to deviate
 % from the power law (cf.~Figure~\ref{fig:24mum_wtheta_linversion}
 %below).}. 
These best-fit models are shown in
Figure~\ref{fig:24mum_wtheta} with blue and red dotted lines.

The spatial correlation lengths $r_0$ were then obtained from the
Limber inversion (Equation (\ref{frm:LE_power})) using the fits to the
empirical redshift distributions of GOODS survey sources, described in
Section~\ref{subsec:24mum_redshift_distribution}. The derived correlation
lengths are $r_0 = 4.98 \pm 0.28 \, h^{-1}\,$Mpc (comoving) for the
low-$z$ ($z_{\rm mean}=0.7$), and $r_0 = 8.04 \pm 0.69 \, h^{-1}\,$Mpc
for the high-$z$ ($z_{\rm mean}=1.7$) sample. Without using a fixed
power-law slope, we obtain $r_{0}=5.07\pm0.34\,h^{-1}\,$Mpc,
$\gamma=1.63\pm0.11$, and $r_{0}=7.99\pm0.75\,h^{-1}\,$Mpc,
$\gamma=1.65\pm0.20$, for the low and high-$z$ subsamples,
respectively.

The uncertainties above include only statistical errors in the
measurement of the angular correlation function. In principle, another
source of uncertainty is the inaccuracies in the models for the
redshift distribution. These are hard to estimate in our case since we
use an empirical fit to the $dN/dz$ observed for the GOODS sources and
any inaccuracies would be related to problems with the GOODS
photometric redshifts.\footnote{We are unaware of such problems, and in
  any case, their discussion is beyond the scope of our work.} The
range of theoretical models for the redshift distribution of
24\,\mum{} sources provides a poor guidance because these models,
still poorly constrained by observations, sometimes give contradictory results \citep{Desai2008, photz2008,
 Franceschini2010}. Qualitatively, if the real $dN/dz$ distribution
for our sources is wider than what we assume, the correlation lengths
should be corrected upward.

As a further check, we re-estimated the correlation lengths for our
high-$z$ subsample using the redshift distribution of the 24\,\mum{}
sources in the COSMOS field \citep{2007ApJS..172...86S,
  2009ApJ...703..222L, 2009ApJ...690.1236I}. The COSMOS survey area is
significantly larger than GOODS ($\approx$2~deg$^{2}$ versus $\approx$0.1~deg$^{2}$) and thus is more representative of our Lockman Hole
region. Unfortunately, there are two problems which prevent us from
using the COSMOS $dN/dz$ as our baseline model. First, the optical and
near-IR data in the COSMOS field are shallower than those in GOODS,
which can affect the $dN/dz$ distribution at high redshifts. Indeed,
7\% of the COSMOS 24\,\mum{} sources with $\f{24}>310\,\mu$Jy have no
redshifts; this is $\approx$20\% of the sources in our high-$z$
bin. Second, there is a significant overdensity of galaxies at $z\sim
1$ in the COSMOS field \citep{2010MNRAS.409..867D}. However, even with
these problems in mind, using the COSMOS-derived $dN/dz$ for the
estimates of $r_{0}$ from the Limber equation provides a useful test
of sensitivity of our results to the assumed shape of the redshift
distribution, possible cosmic variance in the GOODS field, etc. We
applied the same color--magnitude criteria to the 24\,\mum{} COSMOS
sources and approximated the redshift distribution for the high-$z$
bin using either a single-Gaussian model as we do for GOODS, or
two-Gaussian model to better fit a component near $z\sim1$. We derive
$r_{0}=7.90\,h^{-1}\,$Mpc and $8.23\,h^{-1}\,$Mpc for these two
$dN/dz$ approximations, respectively; these values are to be compared
with $r_{0}=8.04\pm0.69\,$Mpc we derive using the GOODS
$dN/dz$. Therefore, this test confirms that the uncertainties in
$r_{0}$ related to the redshift distribution of sources are small
compared to the purely statistical uncertainties.

In what follows, we use the derived correlation lengths for the
24\,\mum{} selected galaxies for estimating the mass range of their
host DM halos through the comparison of our measurements with
the clustering properties of DM halos from the
\emph{Bolshoi} cosmological simulation \citep{Klypin2010}.

%-----------------------------------------------------------------------------
\section{Properties of dark matter halos hosting $24\,\mum$ selected galaxies}
\label{sec:24mum_halo}
%-----------------------------------------------------------------------------

\subsection{Galaxy Population Model}
\label{subsec:galaxy_halo}

Several methods can be used to connect a population of galaxies with
that of their host DM halos \citep[see, e.g.,][and references
therein]{Guo2010}. Here, we use the clustering properties, assuming
that the mass scale of the DM halos hosting the galaxies can
be established by requiring that the observed correlation function of
galaxies selected above a luminosity threshold matches the correlation
function of DM halos selected above a certain mass limit
\citep{2004ApJ...609...35K, Conroy2006} .

To compute the correlation function of the DM halos, we used
the outputs of the \emph{Bolshoi} cosmological simulation for
redshifts ranging from 0.5 to 2.5 with a step size of $\Delta z=0.5$.
The \textit{Bolshoi} simulation, described in \cite{Klypin2010}, is a
high-resolution and large-volume run performed with the WMAP5 and
WMAP7 cosmological parameters $\Omega_{\rm M}=0.27$, $h=0.7$, and
$\sigma_{8}=0.82$ \citep{2009ApJS..180..330K, 2011ApJS..192...18K}. The
simulation contained $2048^3 \approx 8$ billion DM particles
in a $250\,h^{-1}\,$Mpc box. The corresponding mass and force
resolutions are $m_{p}=1.35\times10^{8}\,h^{-1}\,M_{\odot}$ (one
particle mass) and $1.0\,h^{-1}\,$kpc (the smallest cell size in
physical coordinates), respectively. The simulation outputs were
recorded at 180 time steps and were analyzed by the halo-finding
algorithm \citep{Klypin1997, 2004ApJ...609...35K, Klypin2010} to
locate gravitationally bound objects and to calculate their
characteristics such as the virial mass $M_{\rm vir}$, virial radius
$R_{\rm vir}$, maximum circular velocity $v_{\rm max}$, etc. The
identified halos are classified into distinct (host, parent) halos
whose centers are not located within any larger virialized systems,
and subhalos (satellites, substructure) which lie within the virial
radius of a larger halo.  The completeness limit for the halo catalogs
derived from the \emph{Bolshoi} outputs is $v_{\rm max}=50$ \kms{} or
$M_{\rm vir}\approx 1.5\times 10^{10}\,h^{-1}\,M_{\odot}$.

As outlined in \cite{1999ApJ...520..437K}, \cite{2005ApJ...618..557N},
and \cite{Conroy2006}, the maximum circular velocity, $v_{\rm max}$,
of a DM halo, rather than its virial mass, is more closely
related to the properties of a galaxy residing in this halo.
Therefore, we ``populated'' the \emph{Bolshoi} simulation with
``galaxies" by putting the ``galaxies" at the centers of all halos and
subhalos selected above a given $\vmax$ threshold (this threshold
value of $\vmax$ is referred to as \Vmin{} hereafter).  The considered
range of \Vmin{} is $130<\Vmin<385\,$\kms{}. The lower velocity limit
is chosen so that the correlation length for such DM halos is below
the $r_{0}$ derived for our low-$z$ subsample of 24\,\mum{}
galaxies. The high velocity limit is chosen to ensure that the
statistics of DM halos is sufficiently good at all output redshifts of
the \emph{Bolshoi} simulation.  We estimated the correlation lengths
for the model galaxy populations by fitting their spatial correlation
functions with a power law at scales $1<r<25\,h^{-1}\,$Mpc.

\begin{figure}
{\centering
\includegraphics[width=1.0\linewidth]{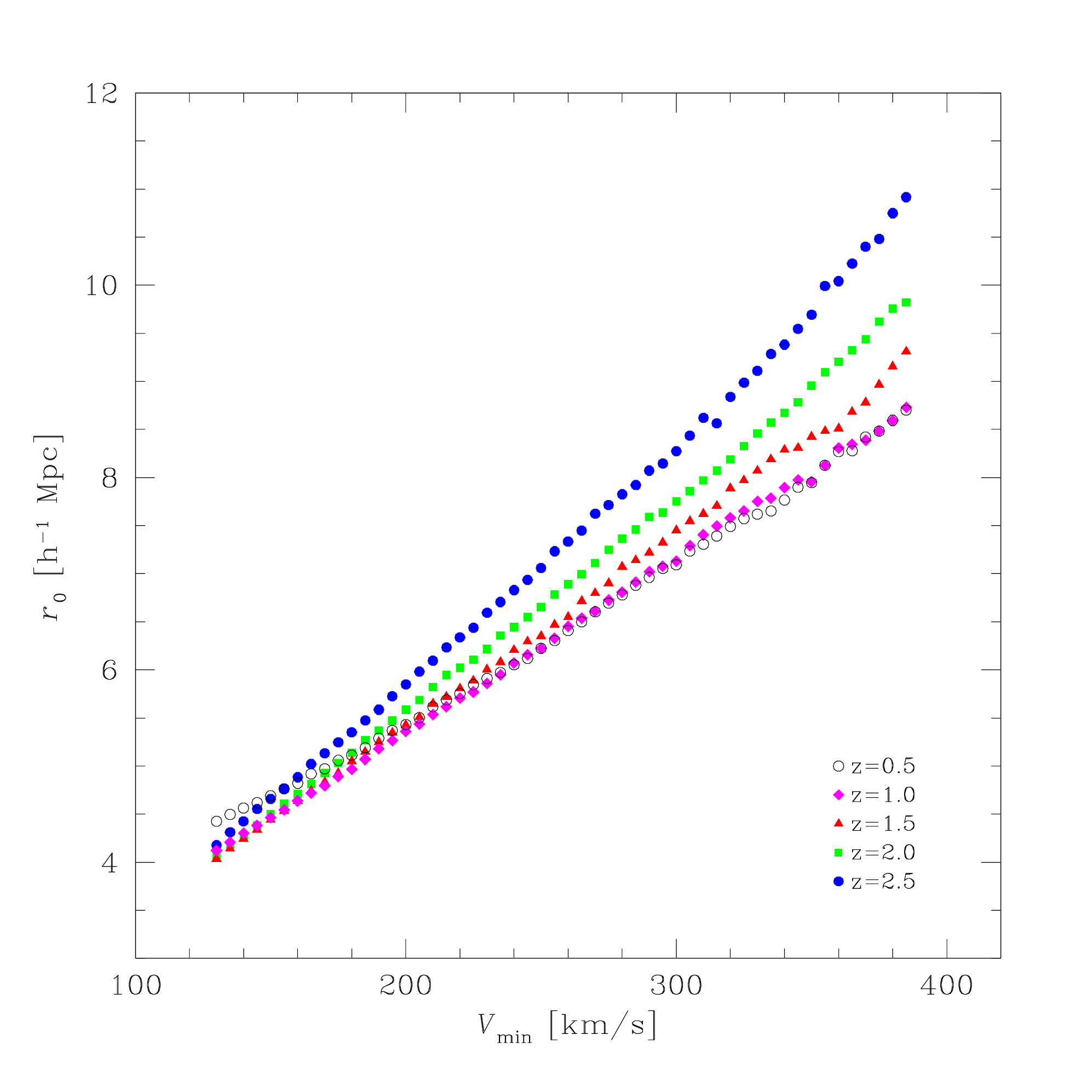}
\par}
\caption{Spatial correlation length of dark matter halos as a
  function of the maximum circular velocity threshold and redshift.}
\label{fig:r0_vmax_bolshoi}
\end{figure}

Figure~\ref{fig:r0_vmax_bolshoi} shows the derived model correlation
lengths for DM halos as a function of \Vmin{} and redshift. Clearly,
the $r_0$ significantly increases with $\Vmin$ (or mass) of the halos
and also changes with redshift. These correlation lengths can be
matched to the observed $r_{0}$ for our samples of 24\,\mum{} selected
galaxies. The redshifts of the simulation outputs do not match exactly
the mean redshifts of our galaxy samples, $z_{\rm mean}=0.7$ and
$z_{\rm mean}=1.7$. However, the trend of the model $r_{0}$ with $z$
for a given \Vmin{} is weak,\footnote{Note that $r_{0}$ as a function
  of \emph{mass} does evolve with redshift, as expected. However, this
  evolution appears to be canceled by the evolution in the $M-\vmax$
  relation and the trend of $r_{0}$ with $M$ at a given redshift.}
and so we can linearly interpolate between the results for the outputs
branching the mean redshifts in the data.

\subsection{Halo Mass and Number Density}
\label{subsec:galaxy_halo_II}

Using these data, each observed value of $r_{0}$ can be matched to the
corresponding \Vmin. The uncertainty intervals for our low- and
high-$z$ subsamples, $r_0 = 4.98 \pm 0.28 \, h^{-1}\,$Mpc and $r_0 =
8.04 \pm 0.69 \, h^{-1}\,$Mpc, respectively, correspond to \Vmin{}
intervals of $\Vmin=172\pm18\,$\kms{} for low-$z$ $24\,\mum$ galaxies
and $\Vmin=322\pm33\,$\kms{} for the high-$z$ subsample with $z_{\rm
mean}=1.7$. 

These velocity thresholds can be easily converted to the corresponding
virial mass limits, $M_{\rm vir}$, using a tight scaling, which
approximately goes as $v_{\rm max}\propto M^{1/3}_{\rm vir}$ \citep[e.g.,][]{Klypin2010}.
This relation is valid for both distinct
halos and subhalos at different redshifts. Fitting the $\vmax-\Mvir$
relation for all halos and subhalos above $\vmax>130\,$\kms{} in the
\emph{Bolshoi} outputs, we obtain the following power-law scalings:
\begin{eqnarray}
\log M_{\rm vir} = 4.60 + 3.25\,\log v_{\rm max}, & \text{for $z=0.5$},   \\
\log M_{\rm vir} = 4.69 + 3.13\,\log v_{\rm max}, & \text{for $z=1.5$},
\label{eq:mass_v_relation}
\end{eqnarray}
where $M_{\rm vir}$ is in units of $h^{-1}\,M_{\sun}$. These results
can be scaled to the mean redshifts of our samples using the expected
redshift evolution of the $\vmax-M_{\rm vir}$ relation, which goes as $M_{\rm vir}\propto
E(z)^{-1}$ for a fixed $\vmax$ \citep{Borgani2009}, where
$E(z)=H(z)/H_{0}$. Using these scalings, we find that the limiting
total mass for the $24\,\mum$ emitting galaxies with $z_{\rm
  mean}=0.7$ is
$\Mtot=(0.7\pm0.2)\times10^{12}\,h^{-1}\,\Msun$\footnote{For
  reference, the Milky Way dark matter halo is estimated to have
  $\vmax=201\,$\kms{} and $\Mtot\sim1.4\times10^{12}\,h^{-1}\,\Msun$
  \citep[e.g.,][]{Guo2010}.} and
$\Mtot=(3.1\pm1.0)\times10^{12}\,h^{-1}\,\Msun$ for our high-$z$
sample.

Having this established mass scale, we can approximately estimate the
fraction of massive DM halos containing 24\,\mum{} emitting galaxies,
even though our sample is not volume-limited. The observed comoving
number density of the galaxies near the mean redshift of the sample
can be estimated as
\begin{eqnarray}
n_{\rm gal} = \frac{dN/dz}{dV/dz} = 1.1\times10^{-3}\,h^3\,{\rm
  Mpc^{-3}}, & \text{$z_{\rm mean}=0.7$},   \\
n_{\rm gal} = 0.12\times10^{-3} \,  h^3\,{\rm Mpc^{-3}}, & \text{$z_{\rm mean}=1.7$},
\label{frm:24mum_number_density}
\end{eqnarray}
where $dV/dz$ is the comoving volume within the survey area. These
values are compared with the number density of halos in the
\emph{Bolshoi} outputs above the derived $\Vmin$ thresholds. For
$z=0.5$, $\vmax>172\,$\kms, we find $n_{\rm halo}= 5.0\times10^{-3} \,
h^3\,{\rm Mpc^{-3}}$, or $n_{\rm halo}\approx5\,n_{\rm gal}$. For
$z=1.5$, $\vmax>322\,$\kms, the corresponding number densities are
$n_{\rm halo}= 0.48\times10^{-3} \, h^3\,{\rm Mpc^{-3}}$ or $n_{\rm
  halo}\approx4\,n_{\rm gal}$.\footnote{The halo number densities at
  the mean redshifts of our samples were determined by the
  interpolation using the closest output redshifts of the
  \emph{Bolshoi} simulation.}  Therefore, we find that similar
fractions, $\sim 20\%$, of DM halos contain galaxies with
$\f{24}>310\,\mu$Jy at both low and high redshifts. This may be simply
a coincidence since the mass and 24\,\mum{} luminosity scales for the
two samples are quite different and so we cannot separate the
luminosity and redshift dependences.

\subsection{Full Limber Modeling of the Observed Angular Correlation Function}
\label{subsec:galaxy_halo_III}

\begin{figure}
{\centering
\includegraphics[width=1.0\linewidth]{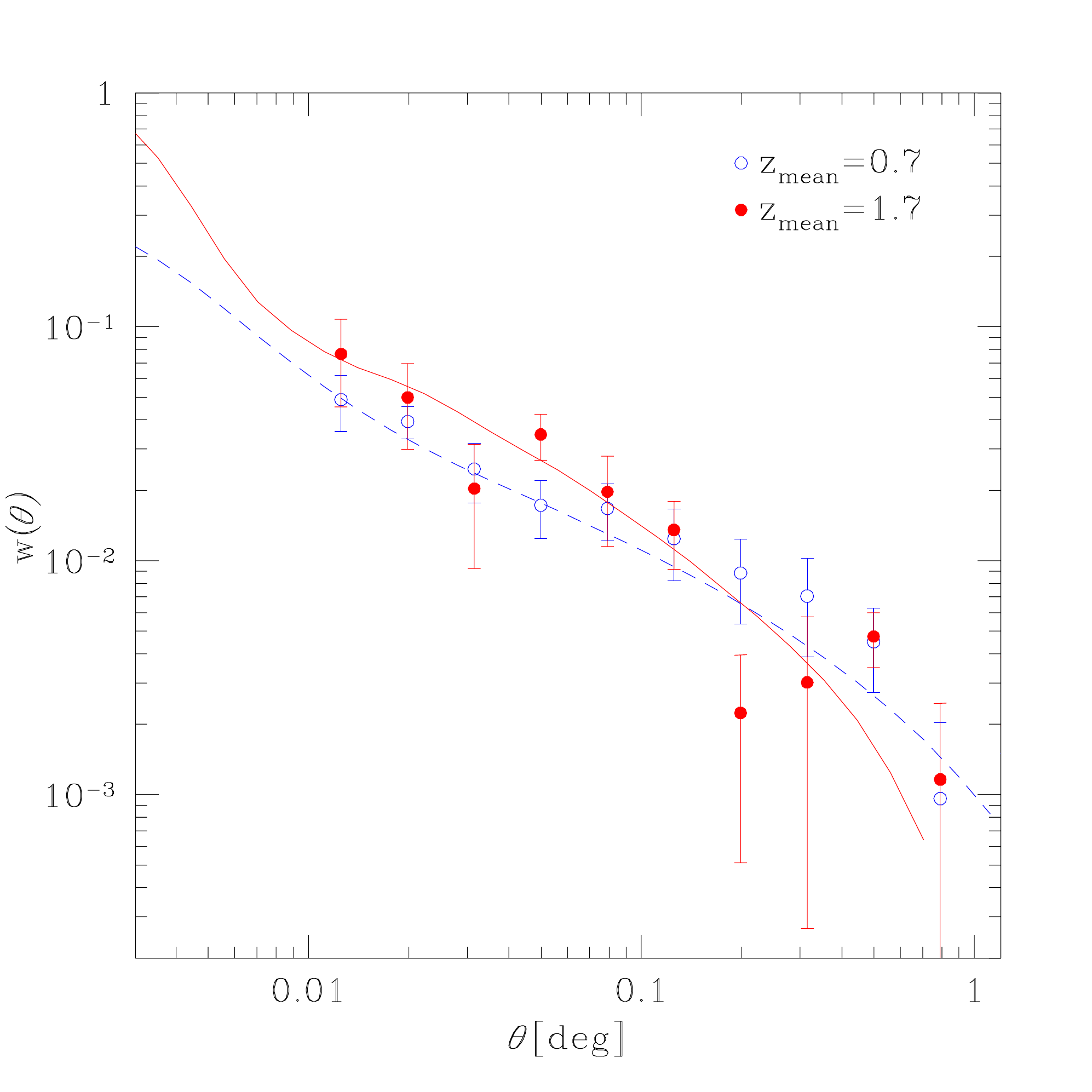}
\par}
\caption{Observed two-point angular correlation function for
low-$z$ (open circles) and high-$z$ (filled circles) samples of the
$24\,\mum$ selected galaxies. The dashed and solid
lines are the angular correlation function models derived from the full
Limber inversion of spatial correlation functions of DM halos with
maximum circular velocities greater than $V_{\rm min}=172\,\,{\rm km
 \,\, s^{-1}}$ and $V_{\rm min}=322 \,\,{\rm km \,\, s^{-1}}$.}
\label{fig:24mum_wtheta_linversion}
\end{figure}

Finally, we test that our analysis based on the power-law
approximation of the observed angular correlation functions provides
unbiased answers even though the correlation function of DM halos
shows clear deviations from the power law at both small and
large scales \citep{2004ApJ...609...35K, Springel2005}. For
this, we compute a full projection of the two-point spatial
correlation function of the \textit{Bolshoi} DM halos for $\vmax>172\,$\kms{} at $z=0.5$ and
$\vmax>322\,$\kms{} at $z=1.5$.\footnote{Note that in calculating the
  projected models, we neglected the redshift evolution of the DM halo
  correlation function within the redshift intervals covered by the
  data. As is clear from Figure~\ref{fig:r0_vmax_bolshoi}, the change
  in the clustering length at our \Vmin{} thresholds is comparable to
  the statistical uncertainties for the $r_{0}$ measurements, so this
  assumption is justified.} The spatial correlation functions, $\xi(r)$, for the halos were
calculated at scales $0<r<50 \, h^{-1}{\rm Mpc}$ in narrow,
$\Delta\log r=0.1 \, h^{-1}{\rm Mpc}$, bins, and then were used in the
full Limber (\citeyear{1953ApJ...117..134L}) transformation:
\begin{equation}
w(\theta) = {\displaystyle{\frac{2}{c}}}\;\frac{\int\limits_{0}^{\infty}dz\,N(z)^2\,H(z)\int\limits_{0}^{\pi_{\rm max}}d\pi\,\xi\left(\!\sqrt{[D_{\rm M}(z)\theta]^2+\pi^2}\right)}{\big[\int\limits_{0}^{\infty}dz\,N(z)\,\big]^2},
\label{frm:LE_full}
\end{equation}
where the functions are the same as in Equation~(\ref{frm:LE_power}), and $\xi(r)=\xi\left(\!\sqrt{[D_{\rm M}(z)\theta]^2+\pi^2}\right)$
is the three-dimensional correlation function under approximation of small angles ($\theta\ll1$ \,[rad]), $\pi$ is the radial separation. 
The results are shown in Figure~\ref{fig:24mum_wtheta_linversion}. The blue
and red data points (open and filled circles, respectively) show the observed angular correlation functions
for the low-$z$ and high-$z$ samples (same as those in
Figure~\ref{fig:24mum_wtheta}), and the lines are the
full projections of the halo correlation functions for the best fit values of \Vmin.

Clearly, the full models fit the data points very well, confirming
that the power-law approximation to the observed $w(\theta)$ yields
accurate estimates of the spatial correlation lengths, $r_0$, and thus
accurate mass scales for the DM halos hosting the 24\,\mum selected
galaxies. At $\theta>0.2$ deg we observed a decline of
the observed correlation functions relative to the power-law
approximations, and this could be related to the behavior of 
the DM halos correlation function at large scales 
\citep[e.g.,][and model curves in Figure~\ref{fig:24mum_wtheta_linversion}]{Springel2005} .

At the opposite end, $\theta<0.01$ deg,
the models show enhancements in the clustering signal relative to the
power-law extrapolation from large radii. These enhancements correspond to the correlation
function of galaxies located within a single parent halo \citep[the so-called
``one-halo'' term,][]{2002PhR...372....1C, 2004ApJ...609...35K}. The measurements of the correlation function at these scales are very interesting because they can be used to determine the location of galaxies in the host DM halos, and thus to constrain their recent merger history \citep[e.g.][]{Porciani2002, 2006ApJ...642...63L, Quadri2008, Cooray2010}.
Unfortunately, the broad PSF of the MIPS instrument does not allow us to make reliable
measurements of the clustering of 24\,\mum{} sources at such small
scales (see discussion~in~Appendix~\ref{sec:ang_fun_comparison}).

%--------------------------------------------------------------------------------------------------------
\section{Comparison with previous measurements}
\label{sec:24mum_comparison}
%--------------------------------------------------------------------------------------------------------

It is important to compare our measurements with the previous studies
of the clustering properties of $24\mum$ selected galaxies. In doing
so, we should keep in mind that direct comparisons with other studies
are difficult because of a wide variety of criteria used for selecting
high-redshift sources. The comparison presented below is done in terms
of the correlation lengths. We do not compare the derived halo masses
because their estimates depend on the assumptions on the cosmological
parameters, power spectrum, and halo occupation models
\citep[e.g.,][]{2008ApJ...679.1192C}, and even the definition used
(e.g., threshold versus mean mass for a population).

We start with low-redshift ($z<1$) samples selected in small areas.
\cite{Gilli2007} presented the correlation function measurements of
the $\f{24}>20\,\mu$Jy galaxies with the mean $z\sim0.8$, detected in
the GOODS fields. They found that the correlation length increases
with the infrared luminosity, reaching for LIRGs ($L_{\rm
  IR}>10^{11}L_{\sun}$) a level of $r_0=5.14\pm0.76\,h^{-1}$\,Mpc. Our
estimate of $r_0$ for the low-\textit{z} subsample ($z_{\rm
  mean}=0.7$) is almost identical to this value. Another study,
focused on the bright $24\mum$ emitting galaxies, was performed
by~\cite{Magliocchetti2008}. The galaxies brighter than
$\f{24}=400\,\mu$Jy detected in the SWIRE XMM-LSS field ($0.7 {\rm
  deg}^2$ used in the analysis) were divided into low-redshift (350
sources at $z_{\rm mean}=0.79$) and high-redshift (210 objects at
$z_{\rm mean}=2.02$) subsamples based on photometric redshifts. The
samples are thus comparable to those selected in our work. The derived
correlation lengths were $5.9^{+1.1}_{-1.3}\,h^{-1}$Mpc and
$11.1^{+2.0}_{-2.4}\,h^{-1}$Mpc for the low and high-$z$ subsamples,
respectively. Within uncertainties, these results are in a reasonable
agreement with our measurements. However, our sample contains a much
larger number of sources and covers a wider area, so we were able to
measure the angular correlation function at larger scales
\citep[probing directly the ``two-halo'' term,
e.g.,][]{2002PhR...372....1C} and significantly reduce the statistical
uncertainties.

Several studies were focused on distant ULIRGs ($z\sim2$) but they
used selection criteria in addition to $24\mum${} flux \citep{Farrah2006, 2007MNRAS.375.1121M,
 Brodwin2008}, therefore their and our results should be compared with caution. For example, \cite{Farrah2006} used a
sample of the ULIRGs with $\f{24}>400\,\mu$Jy which also 
had a spectral peak in the $4.5\mum$ and $5.8\mum$ IRAC bands,
corresponding to the redshifted stellar $1.6\mum$ peak. The
$4.5\mum$ peak sources were estimated to be at $1.5<z<2.0$; their derived
correlation length was $r_0=9.40\pm2.24\,h^{-1}\,$Mpc. The $5.8\mum$ peak
sources are at $2<z<3$ and their angular clustering corresponded to
the correlation length of $r_0=14.40\pm1.99\,h^{-1}\,$Mpc.  The Farrah et al.\ $r_0$ for the 24\mum+4.5\mum{} peak sample is higher than (but consistent within the errors) our value for the high-$z$ sample. We note that their results \cite[as well as those of][]
{Magliocchetti2008} are dominated by the angular clustering measurements at small scales, and thus can be biased if one uses a  power-law fit for the angular correlation function \citep{2004ApJ...609...35K, Quadri2007}.
In another work, a sample of dust obscured galaxies \citep[``DOGs'';][]
{2008ApJ...677..943D} was selected. DOGs are mid-IR luminous
($\f{24}>300\,\mu$Jy) and optically faint ($R-[24]>14$)
galaxies estimated to be at $z\sim 2$. Their measured correlation
length is $7.4^{+1.27}_{-0.84}\,h^{-1}\,$Mpc \citep{Brodwin2008}, similar to our value. 

Models of galaxy formation suggest that DOGs and submillimeter 
galaxies \citep[``SMGs'';][]{2002PhR...369..111B} form by mergers of massive ($M_{\rm
  tot}\sim10^{12-13}\,h^{-1}M_{\odot}$) galaxies \citep[see][and references therein]{2010MNRAS.407.1701N}
and may represent different phases in the evolution of a merging
system. It would be interesting to compare the clustering of SMGs and
other classes of ULIRGs, but, unfortunately, the present estimates
of the SMG correlation length is too uncertain
\citep{2004ApJ...611..725B, 2006MNRAS.370.1057S, 2009ApJ...707.1201W,
2009ApJ...707.1766V, 2010A&A...518L..11M, Cooray2010, 2011Natur.470..510A}.
The best available measurements for submillimeter sources 
with redshifts close to our high-$z$ subsample have been presented in \cite{Cooray2010}.
The authors reported a clustering strength of $r_0=3.15\pm0.35\,h^{-1}$Mpc
and $r_0=4.41\pm0.49\,h^{-1}$Mpc for the HerMes-\textit{Herschel} sources detected
down to the 30 mJy at $250\mum$ and $500\mum$. %, respectively. 
The mean redshift of the samples are $z_{\rm mean}^{250}\approx2.1$ and
$z_{\rm mean}^{500}\approx2.6$.
It is unlikely that these sources are directly related to our $24\mum$ selected galaxies
because of very different values of the inferred correlation lengths.

\section{Conclusions}

We presented an analysis of the clustering properties of $24\mum$ emitting
($\f{24}>310\,\mu$Jy) galaxies detected in Lockman Hole---one of the largest fields in the \emph{Spitzer}/SWIRE survey.  The
large number of sources ($\sim 20,000$) and the size of the field
allowed us to detect the clustering signal with high level of
significance and probe large angular scales. Due to the lack of direct
redshift measurements for the objects in the Lockman Hole sample, we
used the optical and near-IR photometric data to separate the sample into
high-redshift and low-redshift galaxies. The selection criteria as
well as the redshift distributions for color-separated subsamples were
empirically established using the catalogs of GOODS $24\mum$ sources
\citep{Rodighiero2009}, whose redshifts were measured spectroscopically
or estimated from multiband photometry. Using a power-law
approximation to the correlation function, we derived the spatial
correlation length $r_0$. We found $r_0
= 4.98 \pm 0.28\, h^{-1}\,$Mpc and $r_0 =8.04 \pm 0.69 \,
h^{-1}\,$Mpc for $z_{\rm mean}=0.7$ and $z_{\rm mean}=1.7$
populations, respectively.  
%From the comparison with the clustering of
%dark matter halos in the \textit{Bolshoi} cosmological simulation \citep{Klypin2010}, we
%conclude that the 24\,\mum-selected starburst galaxies are located
%within dark matted halos with masses $\Mtot\gtrsim7\times10^{11}\,h^{-1}\,\Msun$
%and $\Mtot\gtrsim3\times10^{12}\,h^{-1}\,\Msun$ at low and high
%redshifts, and populate $\sim20\%$ of such halos.

The estimated infrared luminosities showed that our
$24\mum$ selected galaxies belong to populations of distant ULIRGs
and local LIRGs. Based on the clustering analysis, we can
conclude that our 
$24\mum$ selected galaxies represent different populations of objects
found in differently sized DM halos, $\Mtot\gtrsim7\times10^{11}\,h^{-1}\,\Msun$
and $\Mtot\gtrsim3\times10^{12}\,h^{-1}\,\Msun$ at low and high
redshifts, respectively. In each case, the $24\mum$ selected galaxies populate $\sim20\%$ of the halos at these mass thresholds.
 Their high level of
mid-IR luminosities may be caused by similar physical processes (e.g.,
triggered by mergers or interactions), but occurring in different
environments. Further information can be obtained by studying in
detail the dependence of clustering properties on the IR luminosity
at each redshift.

%-------------------------------------------------------------------------------------------------------------------------------

\acknowledgements
We are grateful to A.~Klypin for letting us use the outputs of the \textit{Bolshoi} cosmological simulations.
We thank C.~Jones for careful reading of the manuscript and useful comments. 
S.S. was supported by the Smithsonian Grand Challenges Consortia. 

\emph{Facility:} \emph{Spitzer}  \\

%each redshift. 

%-------------------------------------------------------------------------------------------------------------------------------

\bibliography{clust24um}

%-------------------------------------------------------------------------------------------------------------------------------

\begin{appendix}

Below, we present a study of stability of the correlation function measurements for $24\mum$ sources
through comparison of different source catalogs in the SWIRE fields.
In particular, we use four largest SWIRE fields (Lockman Hole, ELAIS-N1, ELAIS-N2, CDFS) and
three catalogs - two versions of the SWIRE team catalogs (produced in 2005 and 2010, respectively) and our own list of sources
extracted from \emph{Spitzer}-MIPS maps using the wavelet decomposition method \citep{Vikhlinin1998}.

\bigskip 
\section{Catalogs of $24 \mum$ Sources}
\label{sec:24mum_sample}

The first data set we used is publicly available catalogs from the
SWIRE Data Release 2 (version 2005).\footnote{Available at
  \url{http://swire.ipac.caltech.edu/swire}.} These catalogs consist
of the optical, IRAC, and MIPS 24\,\mum{} information merged into a
single table for sources detected in the IRAC 3.6 and 4.5\,\mum{}
bands above pre-defined SNR thresholds. Source detection in the MIPS
data was carried out using \textit{SExtractor}
\citep{1996A&AS..117..393B}. The estimated completeness threshold is
$\sim 400\,\mu$Jy in all fields. For the clustering analysis, we
selected all $24\,\mum$ sources above this flux threshold. To
eliminate Galactic stars (see Section~\ref{subsubsec:star_gal_sep}), we
cross-correlated this set of 24\,\mum{} sources with the objects in
the 2MASS survey using a matching radius of $2.5''$. Hereinafter, we
refer to these source catalogs (with stars eliminated) as the
``2005-catalog" or ``v.2005".

The second set of catalogs is based on the SWIRE Final Data Release
(J. A. Surace et al. in preparation), a re-reduction of both the IRAC and
MIPS datasets reaching a fainter flux limit. Ancillary
multi-wavelength photometry from the FUV to the NIR was compiled for
sources detected at either $3.6\,\mum$ or $4.5\,\mum$ into the
so-called Data Fusion (M. Vaccari et al., in preparation). For the
IRAC images, the source detection was again done using
\emph{SExtractor}, while the \emph{MOPEX}/\emph{APEX} package
\citep{2005PASP..117.1113M} was used for MIPS data. The
\emph{MOPEX}/\emph{APEX} package was specifically optimized for
detection of point-like sources in crowded fields, and its application
results in a significant improvement in the completeness limit for
MIPS data, which can be as low as $\sim 200\,\mu$Jy (see below). The
completeness of the IRAC detections was also improved compared to the
previous data release.  The initial IRAC source was associated with
the data from other catalogs (e.g., the 2MASS~PSC) using a matching
radius of $2.5''$.  In order to avoid source confusion and false
identification in the $24\,\mum$ band, Vaccari~et~al.\ matched
$24\,\mum$ and IRAC sources within the same radius of $2.5''$. For our
analysis, we used all these $24\,\mum$ sources, and the selected
sample is referred to as the ``2010-catalogs" or ``v.2010".

Another significant difference between the 2005- and 2010-catalogs is
in the methods of flux measurements for the MIPS sources. The 2005
data release used the aperture photometry with a set of apertures
$7.5''-15''$ radius, which contained $60\%-85\%$ of the total flux, and 
applying suitable aperture corrections as determined by the MIPS
instrument team. The
\emph{MOPEX}/\emph{APEX} package yields the total fluxes provided by the PSF
fitting. This is significant in our case because the aperture and PSF
fitting photometry have different problems in dealing with the close
source pairs, which can produce different results for the small-scale
clustering.

Because, as we show below, neither the 2010- nor 2005- catalogs are
completely free of problems, we produced our own list of MIPS-detected
sources (see Section~\ref{subsec:wavelet} for details).
This third data set is referred to as the ``A1-catalog" below.

All $24 \mum$-IRAC catalogs were cross correlated with the 2MASS  survey \citep{Skrutskie2006}
in order to identify and remove foreground stars using \cite{Shupe2008} criterion and to built 
region masks (Section~\ref{subsubsec:mask_construction}). It appeared that in general Galactic stars comprise $\sim 2\%$ 
to the total number of sources detected in the $24\mum$-IRAC bands of SWIRE images.

\section{Limiting Fluxes for Individual Catalogs}
\label{sec:flux_limits}

For a proper comparison of the angular correlation function between
different versions of the source catalogs and different fields, we
have to make sure that the sources are selected above a flux which
exceeds a completeness limit for each field/catalog. Ideally, a
completeness limit is a flux threshold above which (nearly) all real
sources are detected and into which (almost) no fainter sources
migrate. The exact completeness limit for the MIPS/SWIRE data can be
established only through Monte Carlo simulations
\citep[e.g.,][]{Shupe2008}. However, we can apply a useful empirical
criterion and identify the sensitivity limit with a point of maximum
in the differential $\log N$ -- $\log S$ distribution observed for each
field/catalog.

\begin{figure*}
\vspace*{-7mm}
{\centering
\includegraphics[width=0.5\linewidth]{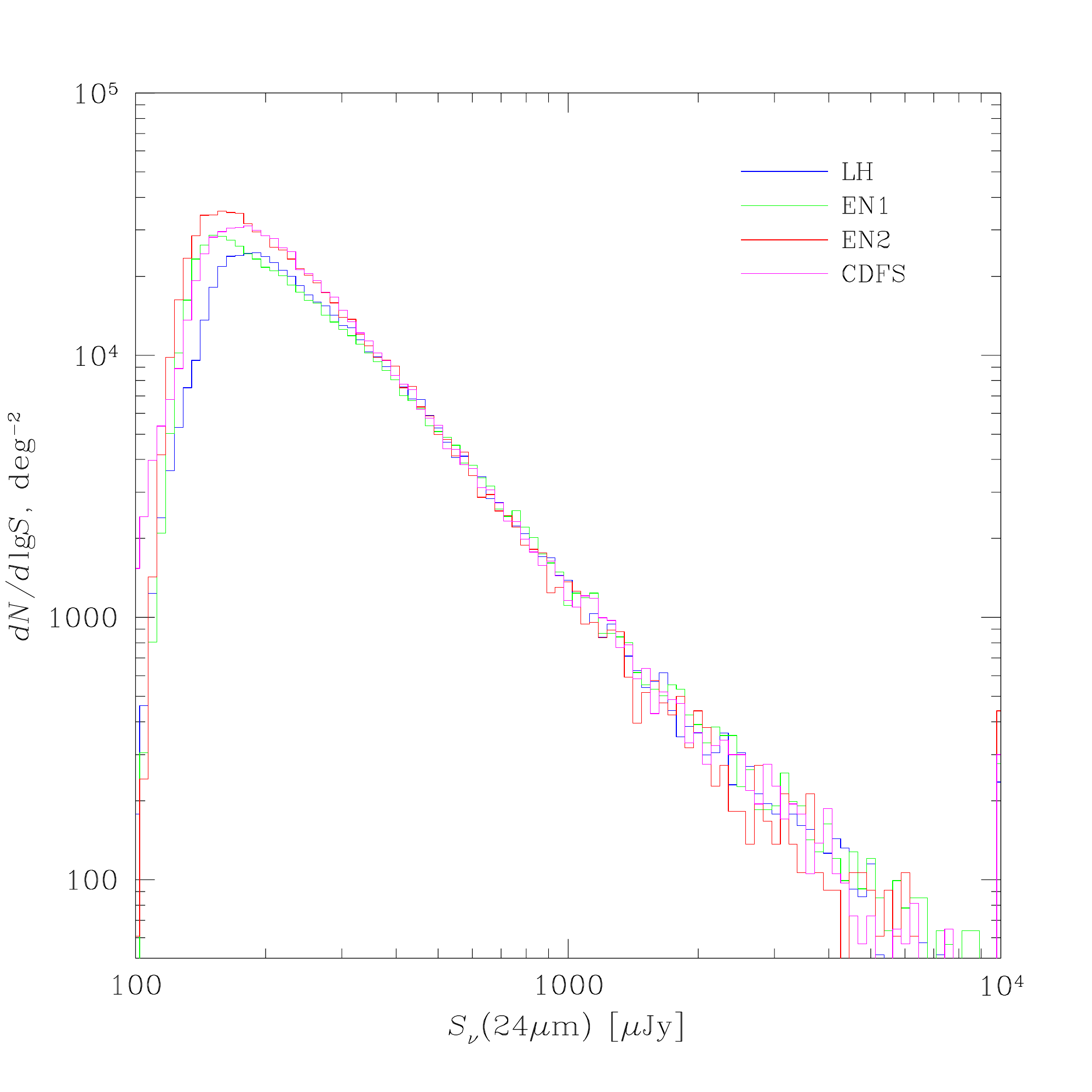}~
\includegraphics[width=0.5\linewidth]{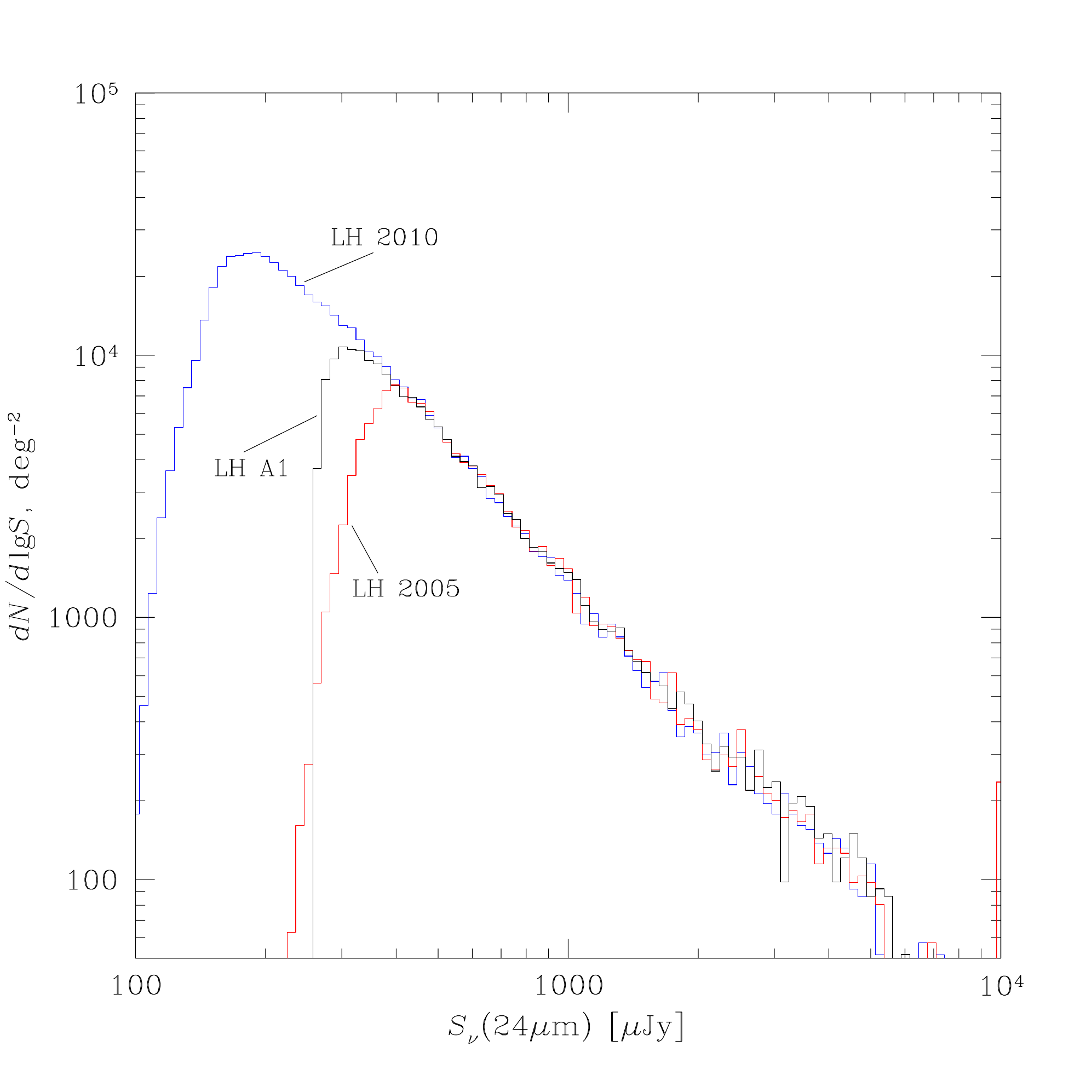}%
\par}
\vspace*{-3mm}
\caption{Number of $24\mum$ sources per square degree per $\log$-flux
  interval plotted vs.~the logarithmic flux. Left: the sources were selected from
  the 2010-catalogs (M. Vaccari et al., in preparation) in the four SWIRE
  fields---Lockman Hole (blue), ELAIS-N1 (green), ELAIS-N2 (red),
  and CDFS (magenta). Galactic stars were masked out and eliminated. Right: the sources were selected from three catalogs in the Lockman Hole. Blue, red, and black lines are for 2010-, 2005- and A1-catalogs, respectively.}
\label{fig:flux_all}
\label{fig:flux_lh_three}
\end{figure*}

In Figure~\ref{fig:flux_all}, we show the number of sources per square
degree and the logarithmic flux bin contained in the 2010-catalogs
for different SWIRE fields. The maxima in the differential $\log N$ --
$\log S$ distribution in all cases are achieved near a flux of $\sim
200\,\mu$Jy. However, there are clear differences in the number counts
of faint sources up to a flux limit of $\f{24}\sim
350\,\mu$Jy. This probably indicates a flux measurement uncertainty of
$\sim 100\,\mu$Jy, which may explain also why the drop in the
differential $\log N$ -- $\log S$ distribution below the point of maximum
is not sharp but extends to $\sim 100\,\mu$Jy. Therefore, based on
examination of the $\log N$ -- $\log S$ distributions, the correlation
functions for the 2010-catalog in different SWIRE fields should be
compared for sources brighter than $350\,\mu$Jy.

\begin{deluxetable*}{ccc}
%\vspace*{-3mm}
\tablecaption{Properties of MIPS SWIRE Fields\label{tab:fields_fluxes}}

\tablehead{
    \colhead{Field} &
    \colhead{$S_{\rm lim}$ ($\mu$Jy)} &
    \colhead{Area ($\rm deg^2$)} 
    }
\startdata
&$S^{2010}=180$&\\
Lockman Hole&$S^{2005}=400$&$8.7$\\
&$S^{\rm A1}=310$&\\
ELAIS-N1&$160$&$7.1$\\
ELAIS-N2&$170$&$3.3$\\
CDFS&$180$&$6.2$\\
ELAIS-S&$S^{\rm A1}=400$&6.3
 \enddata
\tablecomments{The limiting fluxes, $S_{\rm lim}$, reported here correspond to the maxima 
in the source count histograms in Figure~\ref{fig:flux_lh_three}.} 
\end{deluxetable*}

In Figure~\ref{fig:flux_lh_three}, we show the source counts for the
three different catalogs in the Lockman Hole field. There is a
striking difference in the sensitivity limits between the 2005 and
2010 versions of the SWIRE team catalogs---the maxima in the
differential $\log N$ -- $\log S$ distributions are at $\f{24}=400$ and
$180\,\mu$Jy, respectively. The sensitivity limit for the A1-catalog
is between these two values, at $\approx 310\,\mu$Jy. Note that the
drop in number counts below the maximum is very sharp for the A1-catalog, indicating a high level of reliability for the flux
measurements. Even though the $\log N$ -- $\log S$ for the 2010-catalog
extends further down, the flux region $\f{24}\lesssim350\,\mu$Jy in
this catalog might be affected by the scatter in the source flux
measurements, as we have just discussed.

The sensitivity limits (the points of maxima in the differential $\log
N$ -- $\log S$ distribution) for different fields and catalogs are
reported in Table~\ref{tab:fields_fluxes} together with the field
areas after applying the stellar mask (see discussion in
Section~\ref{subsubsec:mask_construction}).  Below, we compare the angular
correlation function computed for different fields/catalogs taking
into account these sensitivity limits.

\section{Comparison of the Angular Correlation Functions}
\label{sec:ang_fun_comparison}  

\begin{figure*}
\vspace*{-3mm}
{\centering
\includegraphics[width=0.5\linewidth]{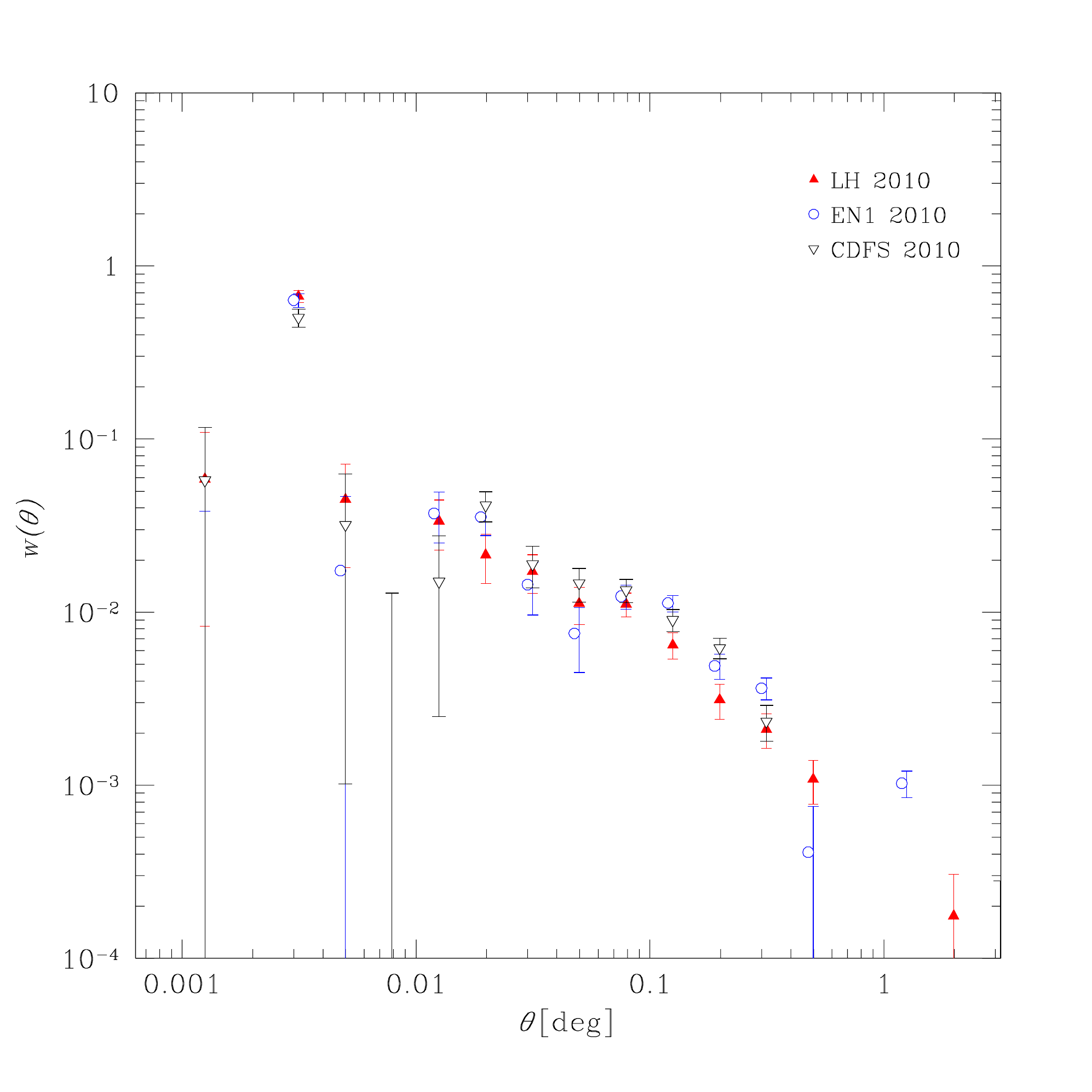}~
\includegraphics[width=0.5\linewidth]{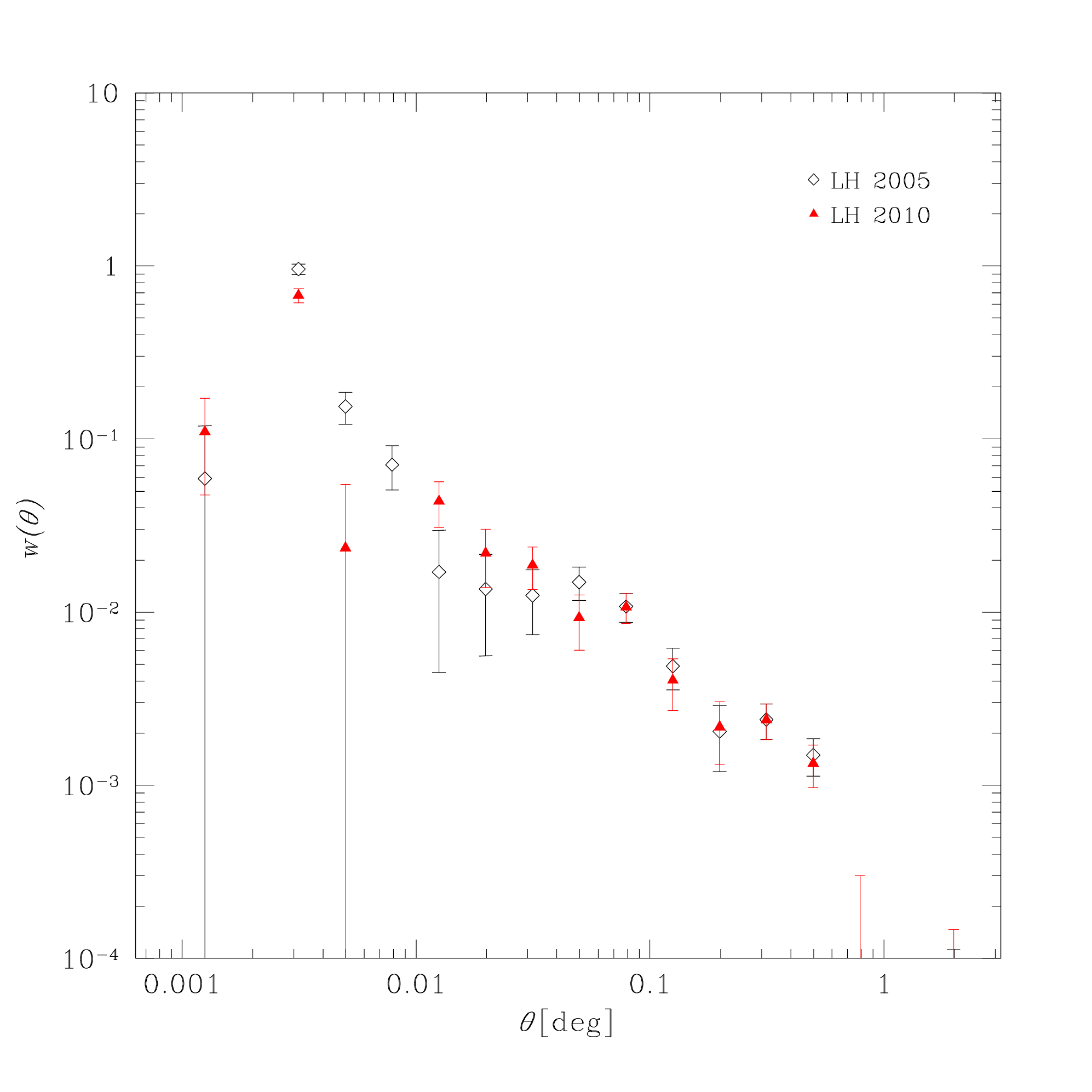}%
\par}
\vspace*{-3mm}
\caption{Left: angular correlation function for sources from
  2010-catalogs with fluxes brighter than $\f{24}=350\,\mu$Jy
  detected in the Lockman Hole (red filled triangles), ELAIS-N1 (blue
  open circles), and CDFS (black open triangles). Right: comparison of the angular correlation
  functions in the Lockman Hole field using the sources from the 2010-
  and 2005-catalogs with $\f{24}>400\,\mu$Jy.}
\label{fig:w_all_300}
\label{fig:w_LH_2005_vs_2010}
\end{figure*}

We start with a comparison of the angular correlation functions,
$w(\theta)$, computed for
different SWIRE fields using the 2010-catalog. As discussed above,
we use a flux threshold of $350\,\mu$Jy. This is the flux above which
the $\log N - \log S$ distributions agree among different fields
(Figure~\ref{fig:flux_all}), and it is higher than the formal
sensitivity limit for the 2010-catalogs. The results are shown in
Figure~\ref{fig:w_all_300} (left). Reassuringly, there is an excellent
agreement between the results in different fields. At the largest
separations, $\sim 1^{\circ}$ and above, the angular correlation
function becomes consistent with zero, but one might expect
distortions at such large scales because they are comparable to the
size of the fields we are using. More relevant to our analysis are the
obvious problems at small scales. There is a drop in the correlation
signal at $0.003^{\circ}<\theta<0.01^{\circ}$, and a strong positive
signal located in a single bin at $\theta\sim 0.003^{\circ}$. As we
discuss below, these distortions are probably related to blending of
nearby sources due to a relatively large size of the MIPS PSF.

Next, we compare the correlation functions for the 2005- and
2010-catalogs above the sensitivity limit for v.2005
($400\,\mu$Jy). The results for the Lockman Hole field are shown in
Figure~\ref{fig:w_LH_2005_vs_2010} (right). There is a good agreement at large
scales ($\theta\gtrsim0.02^{\circ}$) but a strong difference at small
scales. While there is a drop in the correlation signal at
$0.003^{\circ}<\theta<0.01^{\circ}$ for the 2010-catalog sources,
there is a strong excess correlation in the same angular range for the
v.2005 sources. The origin of the discrepancy is probably not because
some real pairs at separations of $\sim 30''$ are missing from the
2010-catalog---it is highly unlikely that this, more sensitive
source list would miss \emph{any} sources brighter that
$400\,\mu$Jy. Rather, we suggest that some of these close pairs arise
spuriously in the 2005-catalog because high fluxes are erroneously
assigned to some faint sources in the vicinity of bright ones \citep[see
also][]{Surace2005}.

\begin{figure*}
\vspace*{-7mm}
\includegraphics[width=0.49\linewidth]{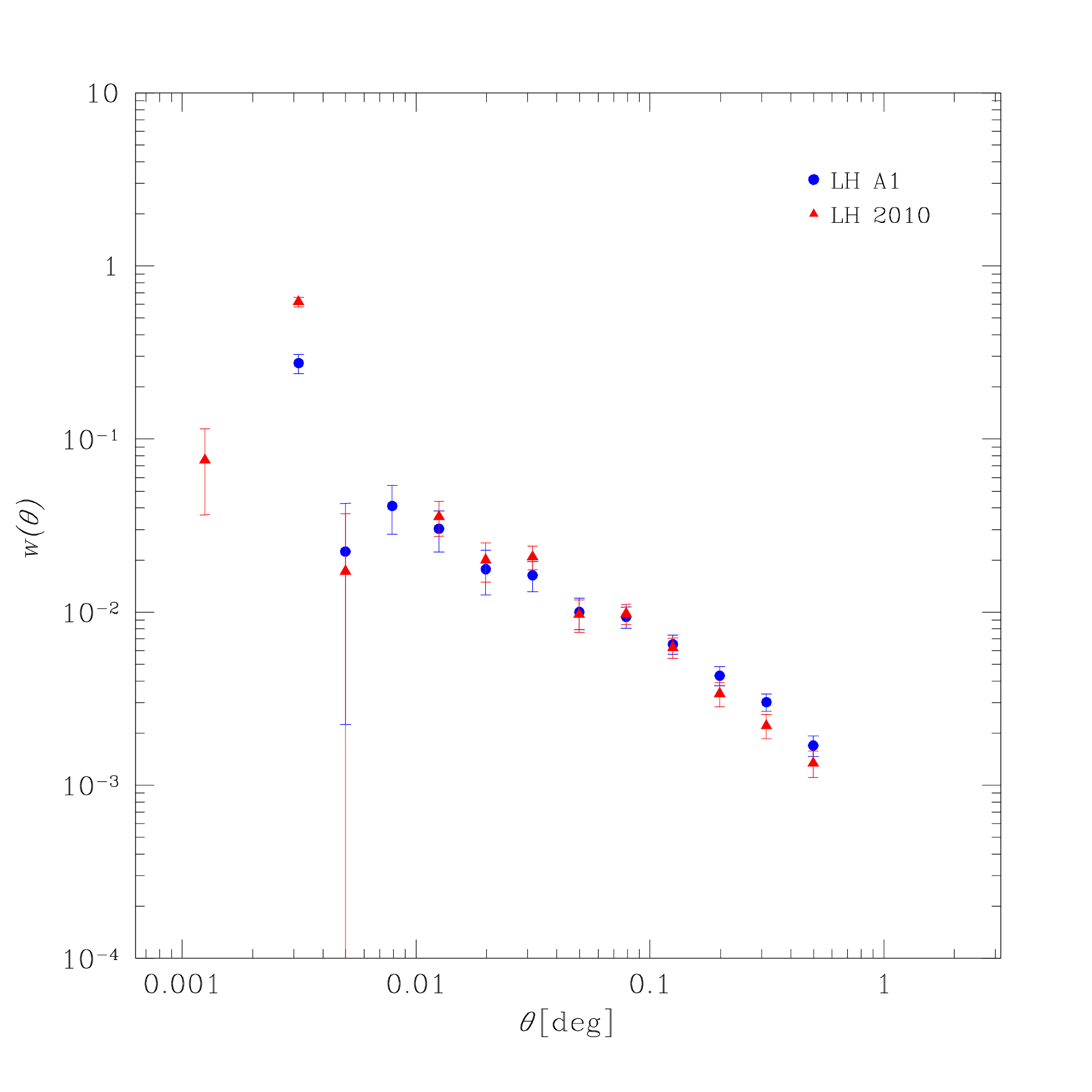}
\hfill
\raisebox{0.055\linewidth}[0pt][0pt]{\includegraphics[height=0.4\linewidth]{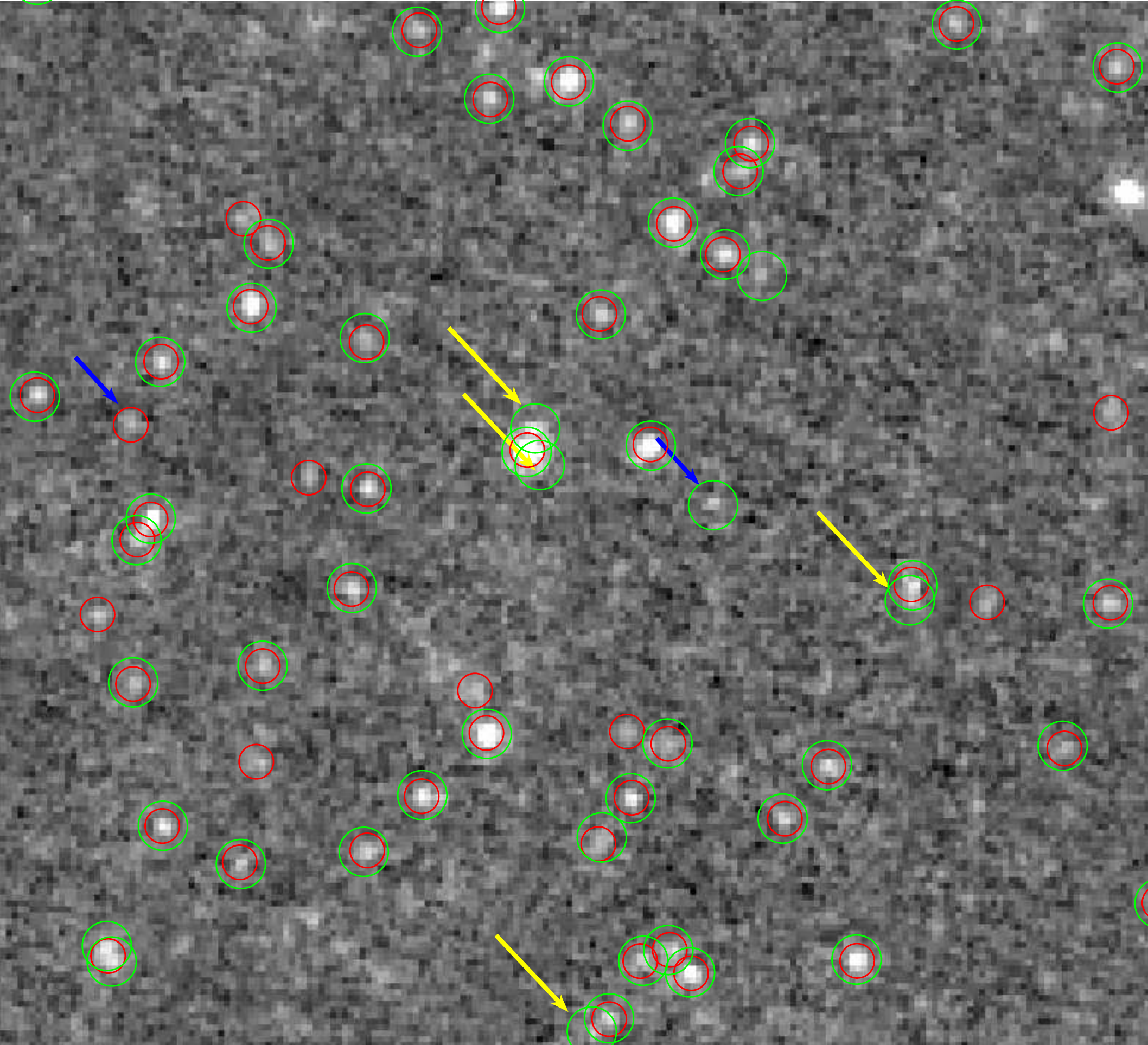}}
\vspace*{-3mm}
\caption{Left: angular correlation function of $24\mum$ sources
  from the 2010 (red filled triangles) and A1 (blue filled circles)
  catalogs using a flux limit of $\f{24}=310\,\mu$Jy in both
  cases. Right: comparison of the bright sources,
  $\f{24}>310\,\mu$Jy, in the 2010- and A1-catalogs (green and
  red circles, respectively) in a subsection of the Lockman Hole
  field. Blue arrows point to \textit{real} detections which are not
  present simultaneously in two catalogs. Yellow arrows indicate
  spurious detections in the 2010-catalog arising in proximity of
  bright/extended sources.}
\label{fig:w_lh_2010_a1}
\end{figure*}

Next, we compare the results for the Lockman Hole field using the
sources from the 2010- and A1-catalogs above a flux threshold of
$\f{24}=310\,\mu$Jy, the sensitivity limit of the A1- catalog. The results are shown in Figure~\ref{fig:w_lh_2010_a1} (left). The
measurements are nearly identical at scales $\theta>0.01^{\circ}$, but
the A1 correlation function shows somewhat weaker small-scale
distortions. This impression is confirmed by cross-examination of the
source detections from both catalogs overlayed on the input MIPS
image (Figure~\ref{fig:w_lh_2010_a1} (right)). Most sources are found in
both catalogs. There are a small number of real sources contained
in one catalog but not the other (examples are marked by blue
arrows) but this is not surprising because the source fluxes are
derived using different methods and so we can expect some
``migration'' across the flux threshold. However, there are some
cases (marked by yellow arrows) where obviously spurious sources are
identified in the 2010-catalog in the vicinity of bright or extended
sources. We believe that these detections are responsible for stronger
small-scale distortions seen in the v.2010 correlation function.

It is clear from the comparisons above that there is a good agreement
in the correlation functions at larger scales, $\theta>0.01^{\circ}$,
when we compare the data for different fields and catalogs above a
common sensitivity threshold. The differences are localized to small
scales and are generally trackable to problems related to blending of
sources in the MIPS images because of a relatively poor angular
resolution of this instrument. These problems are not surprising. The
MIPS PSF has an FWHM of $\approx 6''$ and so the sources become
resolvable only when they are separated by $\sim 10''\approx
0.003^{\circ}$. The MIPS PSF has wide wings---nearly 30\% of the
source flux is scattered outside the $8''$ (radius)
aperture. Therefore, there should be a substantial ``cross-talk'' in
the flux measurements for sources separated by $\sim 15''$ (and up to
$30''$ depending on a source extraction algorithm). In any
case, it appears that the angular correlation function measurements
for the MIPS 24\mum{} sources are not reliable at
$\theta<0.01^{\circ}$, and it is best to restrict the analysis to
larger scales. This is not a problem since our main goal is to measure
the correlation length and the mass scale for the DM halos
hosting the 24\mum{} sources, as these parameters are mainly
constrained by the angular correlation observed near
$\theta=0.1^{\circ}$ (Section~\ref{sec:24mum_wtheta}). 
However, it would be interesting to
put constrains on the location of star-forming galaxies within their
DM halos, which is determined by the shape of the correlation function
at small scales \citep[e.g.,][]{2002PhR...372....1C, 2004ApJ...609...35K}
and thus is not accessible for us.

\begin{figure*}
\vspace*{-7mm}
\includegraphics[width=0.49\linewidth]{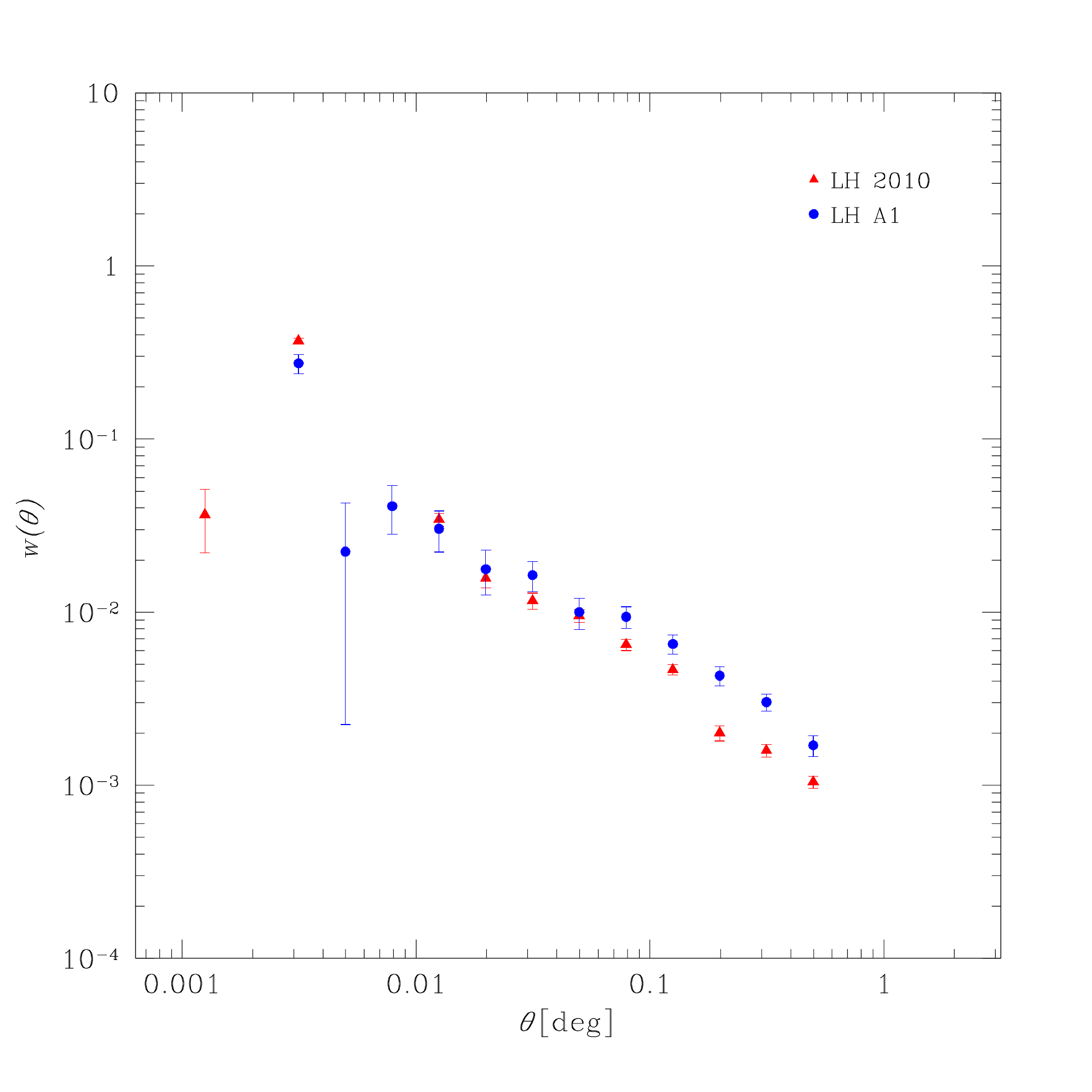}
\hfill
\raisebox{0.055\linewidth}[0pt][0pt]{\includegraphics[height=0.4\linewidth]{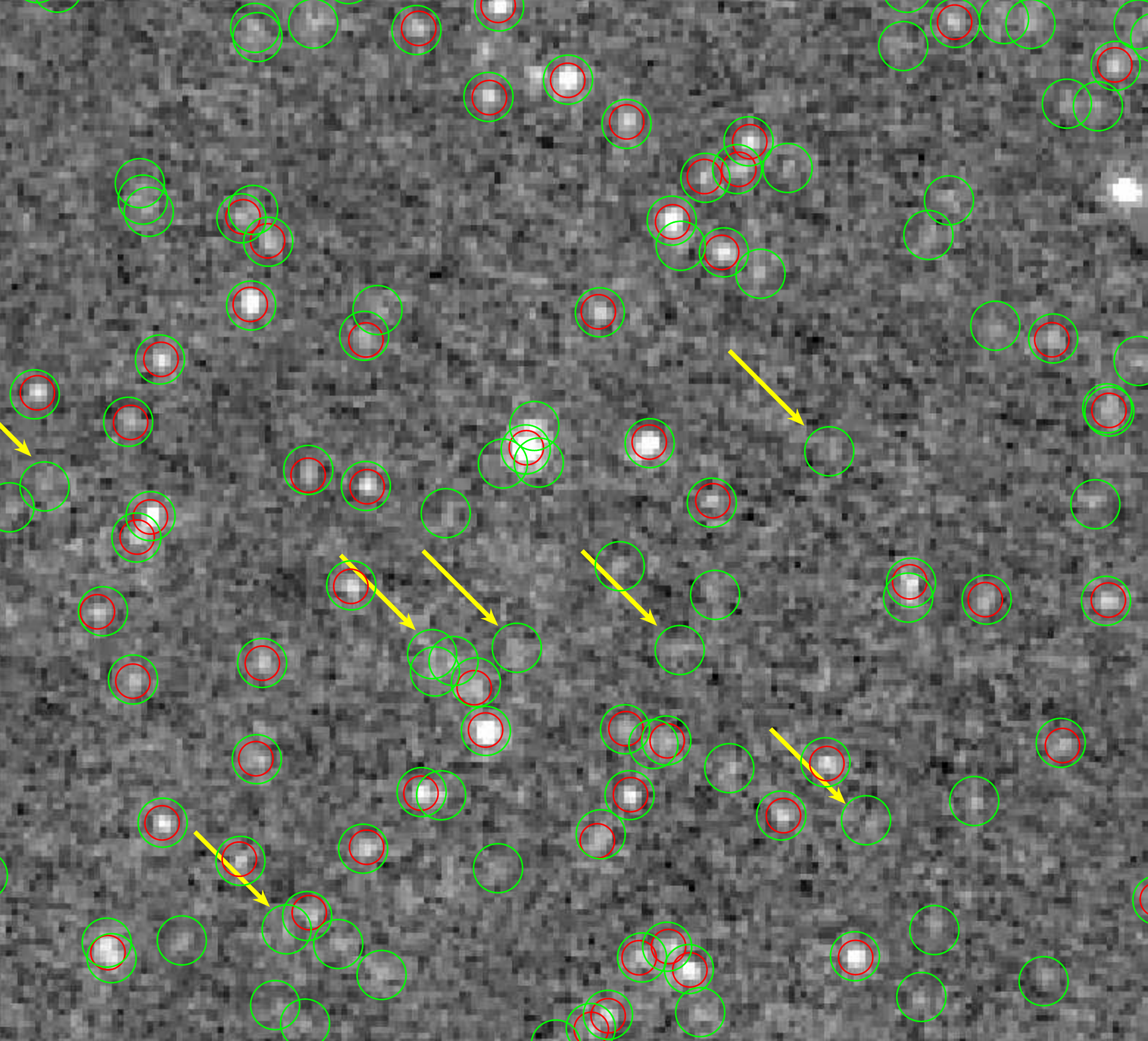}}
\vspace*{-3mm}
\caption{Same as Figure~\ref{fig:w_lh_2010_a1} but the
  2010-catalog sources are selected above a flux limit of
  $\f{24}=180\,\mu$Jy. In the right panel, yellow arrows
  point to the faint sources in the 2010-catalog for which the flux
  measurements are significantly affected by the large-scale
  background variations.}
\label{fig:w_lh_2010_a1_faint}
\end{figure*}

Even though the A1-catalog appears to perform better for the smallest
separations above its flux threshold, $\f{24}=310\,\mu$Jy, the
difference is rather small. The 2010-catalog, on the other hand,
extends to significantly fainter fluxes, and so the question is, can
we use these fainter sources to improve the statistics in the
correlation function measurements? The comparison of the angular
correlation function measurements in the Lockman Hole field for the A1-
and 2010-catalogs above their respective flux limits of 310 and
$180\,\mu$Jy is shown in
Figure~\ref{fig:w_lh_2010_a1_faint} (left). Unfortunately, there are
systematic deviations for the 2010 sources at angular scales
$0.2^{\circ}-0.5^{\circ}$ (recall that the results for the two
catalogs were an excellent agreement for a common flux threshold of
$310\,\mu$Jy, see Figure~\ref{fig:w_lh_2010_a1}). The difference on
these scales cannot be attributed to the edge effects---the size of
the MIPS field in the Lockman Hole region is $\sim4.6\times1.9$
  deg.  Rather, we believe that this difference can be traced to how
the large-scale structures in the MIPS background affect the flux
measurements for fainter sources in the 2010-catalog. Examination of
the MIPS image shows that, indeed, for a significant number of sources
(some marked by yellow arrows in Figure~\ref{fig:w_lh_2010_a1_faint} (right)),
the flux above $180\,\mu$Jy is assigned spuriously, and many such
sources appear on top of larger-scale background structures. These are
likely \emph{real} sources because by construction of the
2010-catalog, they have IRAC counterparts. It is also possible that
these sources are suitable for measurements of the luminosity function
or similar studies because an approximately equal number of objects
``migrate'' below $180\,\mu$Jy in those regions with the negative
residual background. However, for clustering studies, these sources
can not be used because they arise on top of spatially correlated
structures and thus can distort the angular correlation function at
intermediate scales.

\begin{figure}
\vspace*{-3mm}
{\centering
\includegraphics[width=0.49\linewidth]{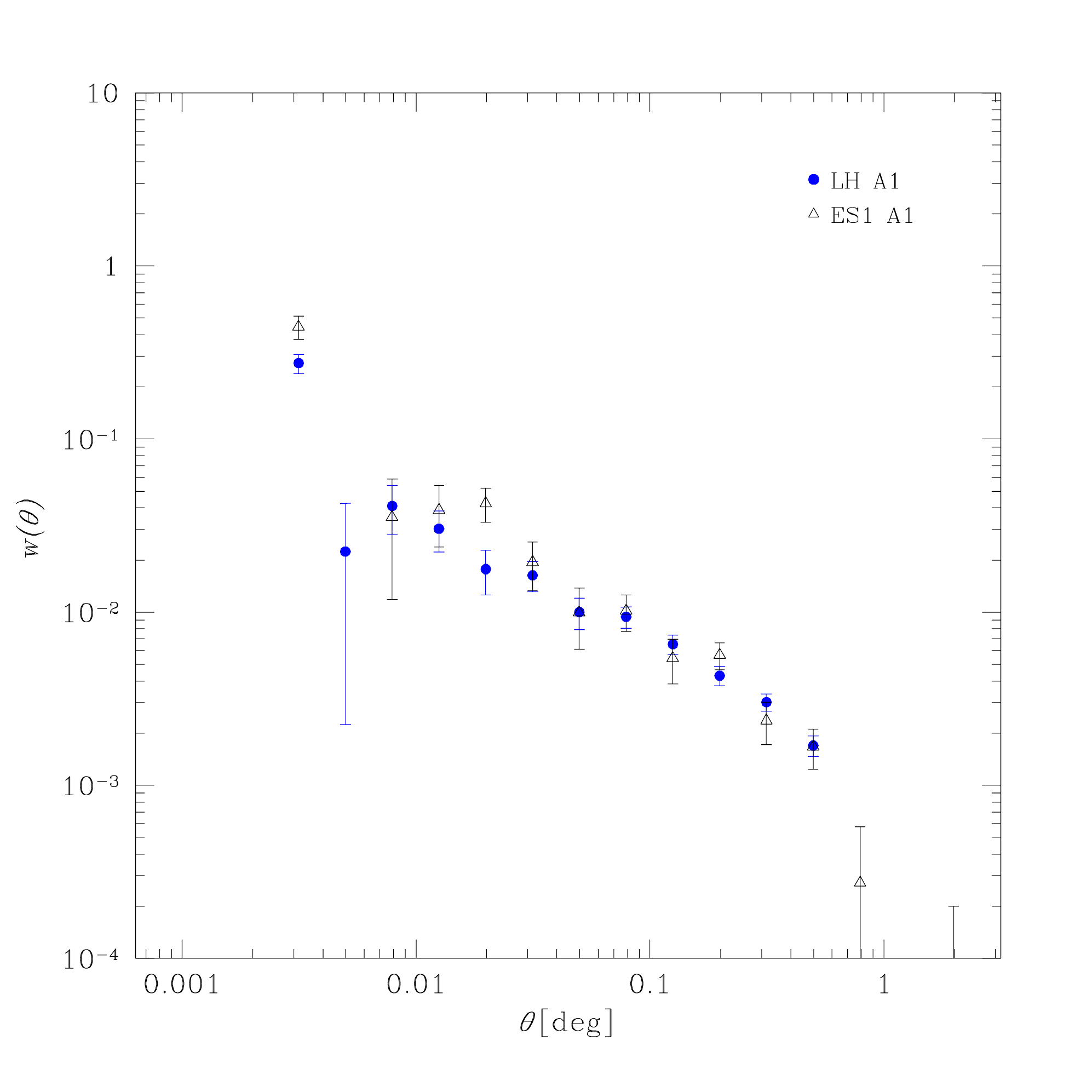}
\par}
\vspace*{-3mm}
\caption{Angular correlation function of $24\mum$ sources from A1- catalogs in Lockman Hole ($\f{24}>310\,\mu$Jy) (blue filled circles) and ELAIS-S1 ($\f{24}>400\,\mu$Jy) (black open triangles).}
\label{fig:w_a1_lh_es1}
\end{figure}

As a final test, we compare the A1-based angular correlation functions
for the Lockman Hole and ELAIS-S1 field
(Figure~\ref{fig:w_a1_lh_es1}). The limiting flux for the A1-catalog
in the ELAIS-S1 field is $\f{24}=400\,\mu$Jy. At all angular
scales, the correlation function computed for sources above this
threshold in the ELAIS-S1 field is in excellent agreement with that
for the Lockman Hole field and $\f{24}>310\,\mu$Jy.

In summary, using our own, completely independent source detection algorithm   
we reproduced the $\log N$ -- $\log S$ at $\f{24}\gtrsim300\mu$Jy and angular correlation function results
at scales $\theta>0.01^{\circ}$ obtained for the 2010-catalog. 
The main analysis presented in
this paper will lead to nearly identical results using either the 2010-
or our A1-catalogs of the 24\,\mum{} sources. The most significant
differences in the measured $w(\theta)$ are localized to
$\theta<0.01^{\circ}$. They can be traced to different treatment of
very crowded regions and zones in the immediate vicinity of bright
sources, where our detection pipeline performs slightly better
(Figure~\ref{fig:w_lh_2010_a1}). On the basis of these considerations,
we choose our A1-catalog in Lockman Hole to investigate clustering
of $24\,\mum$ selected galaxies (Section~\ref{sec:24mum_wtheta}).

\end{appendix}

\end{document}